\documentclass[pra,twocolumn,superscriptaddress,amsmath,amssymb,floatfix]{revtex4}

\usepackage[dvips]{graphicx}
\usepackage{bm}
\usepackage{bbold}
\usepackage{verbatim}

\newcommand{\bra}[1]{\langle #1 |}
\newcommand{\ket}[1]{| #1 \rangle}
\newcommand{\braket}[2]{\left\langle #1 \right|\left. #2 \right\rangle}
\newcommand{\ketbra}[2]{\left| #1 \right\rangle \left\langle #2 \right|}

\newcommand\R{{\mathrm{I\!R}}}

\newcommand{\be}{\begin{equation}}
\newcommand{\ee}{\end{equation}}
\newcommand{\ben}{\begin{equation*}}
\newcommand{\een}{\end{equation*}}
\newcommand{\baen}{\begin{eqnarray*}}
\newcommand{\eaen}{\end{eqnarray*}}
\newcommand{\bae}{\begin{eqnarray}}
\newcommand{\eae}{\end{eqnarray}}

\begin{document}

\title{Quantum phase transitions and quantum fidelity\\
in free fermion graphs}
\author{Marco Cozzini}
\affiliation{Institute for Scientific Interchange, Villa Gualino, Viale
Settimio Severo 65, I-10133 Torino, Italy}
\affiliation{Dipartimento di Fisica, Politecnico di Torino, Corso Duca degli
Abruzzi 24, I-10129 Torino, Italy}
\author{Paolo Giorda}
\affiliation{Institute for Scientific Interchange, Villa Gualino, Viale
Settimio Severo 65, I-10133 Torino, Italy}
\author{Paolo Zanardi}
\affiliation{Institute for Scientific Interchange, Villa Gualino, Viale
Settimio Severo 65, I-10133 Torino, Italy}

\date{\today}

\begin{abstract}
In this paper we analyze the ground state phase diagram of a class of fermionic
Hamiltonians by looking at the fidelity of ground states corresponding to
slightly different Hamiltonian parameters.
The Hamiltonians under investigation can be considered as the variable range
generalization of the fermionic Hamiltonian obtained by the Jordan-Wigner
transformation of the XY spin-chain in a transverse magnetic field.
Under periodic boundary conditions, the matrices of the
problem become circulant and the models are exactly solvable.
Their free-ends counterparts are instead analyzed numerically.
In particular, we focus on the long range model corresponding to a fully
connected directed graph, providing asymptotic results in the thermodynamic
limit, as well as the finite-size scaling analysis of the second order quantum
phase transitions of the system.
A strict relation between fidelity and single particle spectrum is
demonstrated, and a peculiar gapful transition due to the long range nature of
the coupling is found.
A comparison between fidelity and another transition marker borrowed from
quantum information i.e., single site entanglement, is also considered.
\end{abstract}

\maketitle

\section{Introduction}

In the last years, several papers have shown that the study of quantum phase
transitions (QPTs) can benefit from the application of concepts borrowed from
quantum information theory.
In particular, the behavior of entanglement at quantum critical points, due to
its connection with quantum correlations, typically presents peculiar features
which can be used as transition markers, thereby providing further insight into
these phenomena \cite{entanglement}.
More recently, QPTs have also been analyzed from the point of view of Berry's
geometric phase \cite{geometric}, a quantity which has in fact been found to
contribute to entanglement entropy \cite{ryu06}.

One of the most recent reasons of interest in quantum phase transitions for
the quantum information community is the relationship between criticality and
decoherence \cite{decoherence}.
Another relevant aspect is connected with adiabatic quantum computing
\cite{aqc}, as the proximity to QPTs could cause the failure of adiabaticity
conditions.
On the other hand, crossing of QPTs could be desirable to transform
easy-to-prepare states into non-trivial states corresponding to the solution
of computationally relevant problems.
Transitions with non-vanishing energy gap, like topological QPTs, could allow
such an implementation of adiabatic algorithms and therefore the achievement
of significant computational speedups \cite{topological}.

In Refs.~\cite{za-pa,za-co-gio} an approach to quantum phase transitions in
terms of ground state fidelity (the overlap modulus) was proposed.
In practice, this consists in analyzing the fidelity between ground states
corresponding to slightly different Hamiltonian parameters.
The intuitive idea underlying this method is that in the proximity of critical
regions a small change of the Hamiltonian parameters can give rise to dramatic
ground state (GS) variations, due to the strong difference of the GS structure
in opposite phases. The overlap between neighboring states is then expected to
decrease abruptly at the phase boundaries.
In Ref.~\cite{za-pa} this idea was successfully applied to the study of the
Dicke and XY models, where a one to one correspondence was found between the
sudden drops of the overlap function and the critical points of the phase
diagram.
In Ref.~\cite{za-co-gio} a more extensive analysis on general free Fermi systems
was presented, providing a description of the fidelity features in terms of the
matrices of coupling constant entering the system Hamiltonian.
In particular, explicit connections between the vanishing of the lowest single
particle energy (the gap) and the fidelity drops were shown. An exemplification
of these concepts in a model corresponding to a totally connected fermionic graph
was also included.

In this article we further develop the general formalism presented in
Ref.~\cite{za-co-gio}, deriving in detail and expanding some of the results
reported there.
In particular, we characterize the fidelity behavior by a proper combination
$h$ of its second derivatives, namely the lowest eigenvalue of the Hessian
matrix.
Furthermore, we describe a class of exactly solvable models which offer the
ideal playground for the study of the relations between ground state fidelity
and QPTs. The corresponding Hamiltonians are the cyclic variable range
generalization of the fermionic Hamiltonian obtained by the Jordan-Wigner
transformation of the XY spin-chain in a transverse magnetic field.
We also consider their free-ends counterparts, which are however analyzed
numerically, as the analytical results available for the cyclic case make
crucially use of the simple properties of the circulant matrices appearing in the
presence of periodic boundary conditions.
We then focus on the long range model corresponding to a fully connected
fermionic graph.
Along with some first order QPTs, two second order QPTs are found, as anticipated
in Ref.~\cite{za-co-gio} for the free-ends case.
One of them is an example of a \textit{gapful} second order QPT, an uncommon
situation also found in topological QPTs.
The existence of a transition in the absence of a vanishing single particle
energy turns out here to be due to the (unphysical) long range nature of the coupling.
In order to ascertain the second order nature of such transition, in the
analytical cyclic case we also study the ground state energy derivatives, finding
explicitly a discontinuity at the phase boundary in the thermodynamic limit
(TDL).
Furthermore, the finite-size scaling behavior typical of second order
transitions is recovered in the function $h$.
Finally, we calculate the single-site entanglement and compare its behavior with
that of fidelity in critical regions.

The paper is organized as follows.
In Sec.~II we review and discuss in more detail the general results on
quadratic fermionic Hamiltonians already presented in Ref.~\cite{za-co-gio}.
In particular, after a brief summary of the
diagonalization procedure of Ref.~\cite{lieb61}, Subsec.~\ref{subsec:GS} is
devoted to the general expression of the GS in these systems,
Subsec.~\ref{subsec:fidelity} to the explicit calculation of the fidelity, and
Subsec.~\ref{subsec:hessian} to fidelity derivatives and the function $h$.
In Sec.~III we introduce the class of variable range models we have developed.
We begin by recalling the fermionic Hamiltonian obtained by the Jordan-Wigner
transformation of the XY spin-chain in a transverse magnetic field, which
motivated the introduction of our models, and then provide the analytical
solution for the variable range cyclic case.
In Sec.~IV the long range fully coordinated system is described in detail,
first providing an exhaustive analytic description of the cyclic graph
(Subsec.~\ref{subsec:cyclic}) and then presenting the numerical results obtained
for the free-ends one (Subsec.~\ref{subsec:free-ends}).
In this section all the formalism developed in Sec.~I for the fidelity approach
is concretely applied to the study of the phase diagram of the system. The
finite-size scaling of $h$ in the neighborhood of second order QPTs and its
asymptotic behavior in the thermodynamic limit are discussed, along with the GS
energy.
Then, in Sec.~V we report our results on single-site entanglement, whose
behavior in critical regions also shows transition signatures.
Finally, we draw our conclusions in Sec.~VI.
Some technical details can be found in the Appendices.

\section{Ground state fidelity in quadratic fermionic Hamiltonians}
\label{sec:model}

The most general quadratic Hamiltonian describing a system of $L$ free
fermionic modes is given by
\begin{equation}
H=\sum_{i,j=1}^L c_i^\dagger A_{ij} c_j +
\frac{1}{2} \sum_{i,j=1}^L(c_i^\dagger B_{ij} c_j^\dagger +\mathrm{h.c.}) \ ,
\label{eq:qfH}
\end{equation}
where $c_i^{}$, $c_i^\dagger$ are the annihilation and creation operators of the
$i$-th mode and the matrices $A$, $B$ are hermitian and skew-symmetric
respectively, due to the hermiticity of $H$ and the anticommutation properties of
the fermionic operators.
In the following, we restrict ourselves to the case where $A$ and $B$ are {\em
real} matrices.
The generality of the system described by Eq.~(\ref{eq:qfH}) can then be
appreciated by representing $A$ and $B$ in terms of the generic (real) matrix
$Z$, i.e., $A\equiv(Z^T+Z)/2$ and $B\equiv(Z^T-Z)/2$.
This allows to identify the space of coupling constants of Eq.~(\ref{eq:qfH})
with the $L^2$-dimensional full matrix algebra $M_L(\R)$.
Correspondingly, as it will be illustrated below, the system's ground state
properties and their changes in the presence of QPTs will be completely
characterized by the matrix $Z$, which is hence of fundamental importance in our
description.

The above Hamiltonian can be rewritten in the more compact form
$H=(\bm\Psi^\dagger C \bm\Psi+\text{Tr}\,A)/2$,
where $\bm\Psi=(c_1^{},\dots,c_L^{},c_1^\dagger,\dots,c_L^\dagger)^T$,
$\bm\Psi^\dagger=(c_1^\dagger,\dots,c_L^\dagger,c_1^{},\dots,c_L^{})$, and
$C=\sigma_z\otimes{}A+i\sigma_y\otimes{}B$.
This notation, also used in Ref.~\cite{rossini06}, will be adopted throughout
this section.

The diagonalization problem for Eq.~(\ref{eq:qfH}) has been solved in
Ref.~\cite{lieb61}. This well known procedure is very powerful, as it involves
operations in the parameter space of dimension $L^2$, instead of in the full
Hilbert space of the system, which has dimension $2^L$.
Below, we derive the results of Ref.~\cite{lieb61} by making use of the general
properties of real canonical transformations, which we briefly recall also in
view of some simple applications in the next Subsection.

\noindent%
\textit{Real canonical transformations} --
We consider the linear transformation
\be
c_j'= \sum_{k=1}^L (u_{jk}^{}c_k^{} + v_{jk}^{}c_k^\dagger) \ ,
\ee
where $u$, $v$ are {\em real} matrices.
Correspondingly, $\bm\Psi'=V\bm\Psi$, where
$V=\openone_2\otimes{}u+\sigma_x\otimes{}v$ and the subscript is used to specify
the dimension of the unit matrix whenever different from $L$.
The transformation is canonical if it preserves the anticommutation relations.
This implies $V$ to be orthogonal, i.e., the canonical conditions
\begin{subequations}
\label{eq:can_cond}
\bae
&&uu^T+vv^T=\openone \ , \\
&&uv^T+(uv^T)^T=\mathbb{0} \ .
\eae
\end{subequations}
The Hamiltonian becomes
$H=({\bm\Psi'}^\dagger{}C'\bm\Psi'+\mathrm{Tr}A)/2$, where
$C'=VCV^T=\sigma_z\otimes{}A'+i\sigma_y\otimes{}B'$.
Apart from the constant energy shift, which is still given by $\mathrm{Tr}A/2$,
the structure of $H$ is then preserved in terms of the primed quantities.
One can thus define $Z'=A'-B'$, therefore finding by straightforward calculations
the simple relation
\be
\label{eq:Z'}
Z'=(u+v)Z(u-v)^T \ .
\ee

\noindent%
\textit{Hamiltonian diagonalization} --
For the diagonalization of Eq.~(\ref{eq:qfH}) one then needs to diagonalize the
$2L\times2L$ symmetric matrix $C$, i.e., $C=V^TC'V$ with $C'$ diagonal.
We note that the eigenvalue matrix $C'$ can always be written in the form
$C'=\sigma_z\otimes\Lambda$, where the $L\times{}L$ diagonal matrix
$\Lambda=\mathrm{diag}(\Lambda_1,\dots,\Lambda_L)$ is positive semi-definite.
Indeed, it is easy to see
\footnote{Recalling that the determinant of a matrix only changes its sign
under the exchange of matrix rows or columns, one can rearrange the four
$L\times{}L$ blocks which build $C-c\openone_{2L}$ so to prove the implication.}
that $\det(C-c\openone_{2L})=0$ imposes $\det(C+c\openone_{2L})=0$ as well, i.e.,
the eigenvalues of the symmetric matrix $C$ appear in pairs of real numbers of
opposite sign.
On the other hand, one also has $C'=\sigma_z\otimes{}A'+i\sigma_y\otimes{}B'$,
so that $A'=\Lambda$, $B'=0$ and hence $Z'=\Lambda$.
From Eq.~(\ref{eq:Z'}), by defining $\Phi\equiv{}u+v$ and $\Psi\equiv{}u-v$, one
finally gets the important equation
\be
\label{eq:PhiZPsi^T}
\Phi{}Z\Psi^T = \Lambda \ ,
\ee
also implying $\Psi{}Z^T\Phi^T=\Lambda$.
Note that, due to the canonical conditions~(\ref{eq:can_cond}) obeyed by $u$ and $v$, the matrices
$\Phi$ and $\Psi$ must be orthogonal.
It is then straightforward to derive the further relations
$\Lambda\Psi=\Phi{}Z$,
$\Lambda\Phi=\Psi{}Z^T$
and
$\Phi{}ZZ^T=\Lambda^2\Phi$,
$\Psi{}Z^TZ=\Lambda^2\Psi$.
In practice, one can for example solve $\Phi{}ZZ^T=\Lambda^2\Phi$ for $\Phi$
and $\Lambda$, and then obtain $\Psi=\Lambda^{-1}\Phi{}Z$.
Whenever $\Lambda_j=0$, the equation $\Lambda\Psi=\Phi{}Z$ cannot be completely
solved for $\Psi$, but one can recover the $j$-th row of $\Psi$ from
$\Lambda\Phi=\Psi{}Z^T$.
Alternatively, one first solves for $\Psi$ and $\Lambda$, and then for $\Phi$.
Clearly, due to Eq.~(\ref{eq:PhiZPsi^T}), $\Phi$ and $\Psi$ are not independent,
so that in general they cannot be calculated by diagonalizing $ZZ^T$ and $Z^TZ$
separately.

In conclusion, after solving Eq.~(\ref{eq:PhiZPsi^T}) for $\Phi$, $\Psi$
orthogonal and $\Lambda$ diagonal, one introduces $g\equiv(\Phi+\Psi)/2$ and
$h\equiv(\Phi-\Psi)/2$, in terms of which the canonically transformed operators
diagonalizing the Hamiltonian are defined by
$\eta_j\equiv\sum_{k=1}^L(g_{jk}^{}c_k^{}+h_{jk}^{}c_k^\dagger)$.
The Hamiltonian finally reads
\be
H = \sum_{j=1}^L\Lambda_j^{}\eta_j^\dagger\eta_j^{}+E_0 \ ,
\ee
where $E_0=\mathrm{Tr}(A-\Lambda)/2$ is the ground state energy and the
$\Lambda_i$'s are found to play the role of single particle energies.

\subsection{Ground State}
\label{subsec:GS}

In this subsection we present the explicit expression of the ground state
of Eq.~(\ref{eq:qfH}), starting from the Ansatz of Ref.~\cite{peschel}.
The Hamiltonian, although not number conserving due to the terms containing the
antisymmetric matrix $B$, preserves the number parity $P_N=(-1)^N$, where
$N=\sum_{j=1}^Lc_j^\dagger{}c_j^{}$ is the fermion number operator.
In formulas, $[H,N]\neq0$ and $[H,P_N]=0$.
Hence, the ground state must be a superposition of states with even or odd
number of fermions exclusively.
In the case of even parity, it can be of the following form \cite{peschel}
\be
|\Psi_{Z}\rangle = \mathcal{N}
\exp\left(\frac{1}{2} \sum_{j,k=1}^L c_j^\dagger G_{jk}^{} c_k^\dagger\right)|0\rangle \ ,
\label{Psi}
\ee
where $\mathcal{N}$ is the normalization factor,
$\ket{0}$ is the fermionic vacuum ($c_i\ket{0}=0$) and
$G$ is a real $L\times L$ antisymmetric matrix. The link
with the above diagonalization procedure can be made  by
imposing the ground state condition $\eta_k \ket{\Psi_Z}=0,\; \forall \; k$
\cite{peschel}.
Then, the matrix $G$ must be the solution of the equation
\be
\label{eq:gG+h}
gG+h=0 \ ,
\ee
where it is worth recalling $g=(\Phi+\Psi)/2$ and $h=(\Phi-\Psi)/2$.
Note that the latter equation is not always solvable, corresponding to cases
where either the Ansatz~(\ref{Psi}) does not hold or the ground state parity is
odd.
We will return on this point below.

The bridge with the picture of the previous subsection can be made more stringent
whenever $\Lambda$ (and hence $Z$) is invertible.
In fact, by using Eq.~(\ref{eq:PhiZPsi^T}) one can write
$g=\Phi(\openone+\Lambda_\Phi^{-1}Z)/2$,
$h=\Phi(\openone-\Lambda_\Phi^{-1}Z)/2$,
with $\Lambda_\Phi \equiv \Phi^{-1}\Lambda\Phi$,
so that if $g$ is invertible one has
\be
\label{eq:G}
G = \frac{T-\openone}{T+\openone} \ ,
\ee
where $G$ is now seen as the {\em Cayley transform} of
$T\equiv\Lambda_\Phi^{-1}Z$ \cite{bathia}.
The fraction symbol is here not ambiguous, as we are dealing with commuting
matrices, $[T-\openone,(T+\openone)^{-1}]=
[T+\openone-2\cdot\openone,(T+\openone)^{-1}]=0$.
The matrix $T\equiv\Lambda_\Phi^{-1}Z$ can be seen as the orthogonal part of the
polar decomposition of $Z$.
The latter is the generalization to matrices of the polar decomposition of a
complex number $z=|z|e^{i\arg{z}}$.
For a generic matrix $M$ one can write $M=PU$, with $U$ unitary
and $P=\sqrt{MM^\dagger}$ positive semidefinite (note that
$\det{M}=pe^{i\theta}$, with $p=|\det{M}|$ and $e^{i\theta}=\det{U}$).
If $M$ is normal, $[M,M^\dagger]=0$, then one can use without ambiguity the
notation $|M|=\sqrt{MM^\dagger}=\sqrt{M^\dagger{}M}$.
In our case, from Eq.~(\ref{eq:PhiZPsi^T}) we immediately obtain the relation
$P^2=ZZ^T=\Lambda_\Phi^2$ and thus $U=T=\Phi^T\Psi$, real.
Note that, from Eq.~(\ref{eq:G}), the orthogonality $T^T=T^{-1}$ readily implies
the antisymmetry $G^T=-G$.
Since $G$, and thus the ground state $\ket{\Psi_Z}$, can be expressed in terms of
$T$, the latter will be of central importance in our analysis. In fact, many of
the properties of the ground state we are interested in can be related to or
derived from the properties of $T$.

Unless $Z$ is not invertible, $T$ is uniquely defined. Note that $\det{Z}\neq0$
implies $0\notin\mathrm{Sp}\Lambda$, i.e., the system is gapful.
In a gapful phase, therefore, $T$ is always well defined and, provided
$T+\openone$ is invertible, $G$ exists and the GS is of the form~(\ref{Psi}).
The inverse Cayley transform then gives $T=(\openone+G)/(\openone-G)$, which
implies $\det{T}=1$. Indeed, the spectrum of the real antisymmetric matrix $G$ is
given by complex conjugate pairs of purely imaginary eigenvalues (with the
exception of an unpaired zero eigenvalue for $L$ odd), so that for each
eigenvalue $1+\lambda_G$ of $\openone+G$ there exists an eigenvalue
$1-\lambda_G^*=1+\lambda_G$ of $\openone-G$.

The determinant of $T$, or equivalently the sign of $\det{Z}$, turns out to
correspond to number parity, i.e., $\det{T}=1$ and $\det{T}=-1$ for ground
states of even and odd parity respectively. We give a detailed proof in
Appendix~\ref{app:parity}. Here it is important to notice that whenever
$-1\in\mathrm{Sp}T$ the matrix $T+\openone$ is not invertible and $G$ cannot be
defined. Equivalently, $g$ is not invertible and Eq.~(\ref{eq:gG+h}) cannot be
solved.
However, with the elementary canonical transformations described in
Appendix~\ref{app:parity}, one can always reduce to a $T'$ such that
$-1\notin\mathrm{Sp}T'$, so that the ground state can be expressed by
Eq.~(\ref{Psi}) in terms of the transformed quantities.
This will be useful in deriving the fidelity formula of
Subsec.~\ref{subsec:fidelity}.
In addition, provided $L$ is even and $\det{T}=1$, one can write the ground
state in a form which is valid independently of the presence of $-1$ in
$\mathrm{Sp}T$ and can hence be considered more general than Eq.~(\ref{Psi}).
This is \cite{za-co-gio}
\begin{equation}
|\Psi_Z\rangle=\otimes_{\nu=1}^{L/2} [\cos(\theta_\nu/2) |00\rangle_{\nu,-\nu} + \sin (\theta_\nu/2)  |11\rangle_{\nu,-\nu}] \ ,
\label{Psi_D}
\end{equation}
where $\{e^{\pm i\theta_\nu}\}_{\nu=1}^{L/2}=\mathrm{Sp}T$ and $|0\rangle_\nu$
($|1\rangle_\nu$) denotes the vacuum (occupied) state of the new pairs of
fermionic modes $\tilde{c}_\nu=c_{2\nu-1}'$, $\tilde{c}_{-\nu}=c_{2\nu}'$, with
$c_j'=R_{jk}^Tc_k$.
The matrix $R$ is the one putting $T$ and $G$ in block diagonal form (see
Appendix~\ref{app:parity}).
In Ref.~\cite{za-co-gio} Eq.~(\ref{Psi_D}) was derived starting from
Eq.~(\ref{Psi}), but it should be clear from the discussion in
Appendix~\ref{app:parity} how to prove it for cases where $G$ cannot be
defined.
One also recognizes when the Ansatz~(\ref{Psi}) is not valid: if
$-1\in\mathrm{Sp}T$ then $\theta_\nu=\pi$ for some $\nu=1,\dots,L/2$ and hence
the corresponding term in the direct product~(\ref{Psi_D}) is simply
$\ket{11}_{\nu,-\nu}$, which can be described by expression~(\ref{Psi}) only in
a limiting sense (the weight of the $\ket{00}_{\nu,-\nu}$ term cannot vanish
unless the normalization $\mathcal{N}$ diverges).

\subsection{Ground state fidelity}
\label{subsec:fidelity}

We are interested in characterizing how the ground state changes when
the set of parameters of Eq.~(\ref{eq:qfH}) are slightly modified, i.e., when
$(A,B)\rightarrow (A+\delta A, B+\delta B)$ or equivalently when
$Z\rightarrow \tilde{Z}=Z+\delta Z$. In order to characterize these changes we
use the the ground state fidelity, which was introduced in Ref.~\cite{za-pa}
and discussed in Ref.~\cite{za-co-gio}:
\be
{\cal F}(Z,\tilde{Z}) \equiv
|\braket{\Psi_{Z}}{\Psi_{\tilde{Z}}}| \ .
\label{over}
\ee
As already pointed out in Ref.~\cite{za-co-gio}, in order to evaluate it
we can resort to the theory of coherent states \cite{perelemov}.
In fact, the GS construction given by Eq.~(\ref{Psi}) coincides with the
construction of a coherent state for the group $SO(2L,\R)$.
In general the construction of a coherent state relative to a
Lie group $\mathcal{G}$  involves the action of an element of a
unitary representation of the group over the so called maximal weight vector
of the Lie algebra $\mathcal{A}$. In our case, we have that the group coincides
with $SO(2L,\R)$ and the relative Lie algebra $so(2L,\R)$ is isomorphic to the
algebra generated by the operators
$\{c_ic_j,c_j^\dagger c_i^\dagger, \frac{1}{2} (c_ic_j^\dagger -c_j^\dagger c_i) \}.$
The group $SO(2L,\R)$ has two unitary irreducible representations over the $2^L$
fermionic Hilbert space that splits, according to these representations into two
irreducible subspaces even and odd, i.e.,
$\mathcal{H}_F=\mathcal{H}^+_F \bigoplus \mathcal{H}^-_F$,
having the basis vectors with an even and odd number of fermions respectively.
If we now use the even representation, we can generate the generic coherent state
by acting with an element $U^+=\exp(A)$, $A\in so(2L,\R)$ upon the maximal weight
vector of the algebra that in this case is the already introduced fermionic
vacuum ($c_i\ket{0}=0$). If we now use $(A)_{ij}=-G_{ij}c_j^\dagger c_i^\dagger$
we have that the generic coherent state can be identified by a skew-symmetric
matrix $G$ (in our case determined by $Z$) and has exactly the form (\ref{Psi}).
This identification allows us to use the general result for the scalar product
(overlap) between fermionic coherent states defined by the antisymmetric
matrices $G$ and $\tilde{G}$, namely \cite{perelemov}
\be
\braket{\Psi_{\tilde{Z}}}{\Psi_{{Z}}}=\frac{ \det(\openone + G^\dagger \tilde G)^{1/2} }
{\det(\openone + G^\dagger G)^{1/4}\det(\openone + \tilde{G}^\dagger \tilde{G})^{1/4}} \ .
\label{overlap}
\ee
We can then explicitly write the ground state fidelity in terms of $T$.
In fact, as described in Ref.~\cite{za-co-gio} one finds
\be
{\mathcal F}(Z,\tilde{Z}) =
\sqrt{\left|\det\frac{T+\tilde{T}}{2}\right|}=
\sqrt{\det\frac{\openone+T^{-1}\tilde{T}}{2}} \ .
\label{overlap-marco}
\ee
It is not difficult to get the latter result by substituting
$G=(T-\openone)/(T+\openone)$ into Eq.~(\ref{overlap}).
Whenever one of the ground states cannot be expressed
directly into the form~(\ref{Psi}), two possibilities are present.
Either the ground states belong to different sectors of fermion number parity
(that is $\det{T}=-\det{\tilde{T}}$ and $\det(T^{-1}\tilde{T})=-1$) and the
fidelity vanishes, as correctly reproduced by Eq.~(\ref{overlap-marco}),
or they belong to the same sector but $-1$ is an eigenvalue of $T$ and/or
$\tilde{T}$.
In the latter case, a suitable canonical transformation can always be found
(similar to those explained in Appendix~\ref{app:parity}) such that (i) $-1$ is
absent from the spectrum of both the transformed matrices $T'$ and $\tilde{T}'$
and (ii) $\det(T'+\tilde{T}')=|\det(T+\tilde{T})|$.
Consequently, one has
$\mathcal{F}=\sqrt{\det[(T'+\tilde{T}')/2]}=\sqrt{|\det[(T+\tilde{T})/2]|}$ and
formula~(\ref{overlap-marco}) is proven for all the possible cases.

We also recall that if $T^{-1}\tilde{T}\in{}SO_L(\R)$ and $L$ is even
then \cite{za-co-gio}
\be
{\mathcal F}(Z,\tilde{Z}) =
\prod_{\nu=1}^{L/2} |\cos(\Theta_\nu/2)| \ ,
\label{overlap-cos}
\ee
where $\mathrm{Sp}(T^{-1}\tilde{T})=\{e^{\pm{}i\Theta_\nu}\}_{\nu=1}^{L/2}$.

\subsection{Fidelity derivatives}
\label{subsec:hessian}

The fidelity function compares ground states at different points in the
parameter space, giving a measure of their `distance'.
Intuitively, in a critical region the ground state changes significantly even
for small variations of the Hamiltonian parameters. In other words, the rate of
orthogonalization diverges in the proximity of QPTs.
It is then reasonable to characterize the degree of criticality by the
rapidity of the ground state variation as a function of the system parameters.
In practice, this is clearly related to derivatives of the fidelity function.

We fix a point $Z$ in the parameter space, with the corresponding ground state
$|\Psi_Z\rangle$. The fidelity between $|\Psi_Z\rangle$ and the GS
$|\Psi_{\tilde{Z}}\rangle$, corresponding to a second point $\tilde{Z}$, can be
regarded as a function of $X=\tilde{Z}-Z$, defined by
$F_Z(X)\equiv\mathcal{F}(Z,\tilde{Z})$.
Clearly, the function $F_Z(X)$ has a maximum for $X=0$, where $F_Z(0)=1$.
Consequently, the first derivatives of $F_Z(X)$ with respect to $X$ must vanish.
In order to analyze the fidelity behavior around $X=0$ one has then to
consider second derivatives, i.e., the Hessian matrix of $F_Z(X)$ calculated at
$X=0$, which we denote by $H_{F_Z}(0)$.
The eigenvalues of the Hessian matrix correspond to the principal curvatures
of the hypersurface generated by the function $F_Z(X)$ at $X=0$.
The eigenvector corresponding to the lowest (highest in modulus) eigenvalue
gives the direction of most rapid change in the GS.
In the proximity of a QPT one expects a (negative) divergence of the smallest
Hessian eigenvalue.
We then use $h(Z)=\min\{\mathrm{Sp}[H_{F_Z}(0)]\}$ as a measure of the degree
of criticality of the point $Z$.
For the two-dimensional parameter space $\mu$-$\gamma$ considered in this paper
one has
$Z=Z(\mu,\gamma)$, $X=X(\delta\mu,\delta\gamma)$, and
\begin{eqnarray}
\label{eq:h(Z)}
h(Z) & = & \frac{1}{2}
\left[\partial_{\delta\mu}^2F_Z+\partial_{\delta\gamma}^2F_Z+\right. \\
& - & \left.\sqrt{(\partial_{\delta\mu}^2F_Z-\partial_{\delta\gamma}^2F_Z)^2+
4(\partial_{\delta\mu}\partial_{\delta\gamma}{}F_Z)^2}\right] \nonumber \ ,
\end{eqnarray}
where all the derivatives are calculated at $\delta\mu=\delta\gamma=0$
(corresponding to $X=0$).

\section{Variable range fermionic XY models}

We begin this section by briefly reviewing some aspects of the traditional XY
spin chain \cite{lieb61} in a transverse magnetic field \cite{barouch}.
In accordance with recent literature, we write the corresponding ferromagnetic
Hamiltonian as
\be
\label{eq:XY}
H = -\sum_j\left[\frac{1+\gamma}{2}\sigma_j^x\sigma_{j+1}^x+
\frac{1-\gamma}{2}\sigma_j^y\sigma_{j+1}^y
+\lambda\sigma_j^z\right] \ ,
\ee
where $\sigma_j^\alpha$ for $\alpha=x,y,z$ are the usual Pauli operators
relative to the $j$-th site, $\gamma$ is the anisotropy parameter, and $\lambda$
defines the strength of the magnetic field in the $z$ direction.
Actually, in the original papers \cite{lieb61,barouch} an antiferromagnetic
spin chain was chosen.
As recalled later, however, this makes no difference for the ground state phase
diagram.

In Eq.~(\ref{eq:XY}) we did not specify the range in the summation signs. Two
possible choices are indeed possible, (i) the cyclic chain with periodic
boundary conditions and (ii) the free-ends chain. We will return on this point
when considering the fermionic version of this Hamiltonian.

In fact, by using the Jordan-Wigner transformation, Eq.~(\ref{eq:XY}) can be
rewritten in terms of the fermionic operators
\be
c_j = e^{i\pi\sum_{k<j}\sigma_k^+\sigma_k^-}\sigma_j^- \ ,
\ee
where $\sigma_j^\pm=(\sigma_j^x\pm{i}\sigma_j^y)/2$ are the Pauli raising ($+$)
and lowering ($-$) operators.
For a free-ends chain with $L$ sites one gets
\be
\label{eq:cXY}
H = -\sum_{j=1}^{L-1}
(c_j^\dagger c_{j+1}^{} + \gamma c_j^\dagger c_{j+1}^\dagger + \mathrm{h.c.})
-\sum_{j=1}^L \lambda(2c_j^\dagger c_j^{}-1) \ ,
\ee
while for the cyclic case additional boundary terms appear, which are however
negligible in the thermodynamic limit \cite{lieb61}.
Clearly, apart from a constant term, Eq.~(\ref{eq:cXY}) is a special case of
Eq.~(\ref{eq:qfH}).

The Hamiltonian~(\ref{eq:cXY}) corresponds to a free-ends chain in terms of the
$c_i$ fermionic operators. Analogously, one can consider the ``$c$-cyclic''
problem, where periodic boundary conditions are applied to the fermionic chain.
In Ref.~\cite{lieb61} the $c$-cyclic problem is shown to be equivalent to the
periodic spin chain in the thermodynamic limit.
Here, we generalize the fermionic XY Hamiltonian by introducing variable range
couplings and studying both the $c$-cyclic and the $c$-free-ends problems.
In the free-ends case, the Hamiltonian with a coupling of range $r$ reads
\be
\label{eq:fevrcXY}
H = -\sum_{j=1}^{L-1}\sum_{k=1}^{\min(r,L-j)}
(c_j^\dagger c_{j+k}^{} + \gamma c_j^\dagger c_{j+k}^\dagger + \mathrm{h.c.})
-\mu\sum_{j=1}^L c_j^\dagger c_j^{} \ ,
\ee
where $\mu$ plays the role of the chemical potential.
For $r=1$ one reduces to the nearest neighbor coupling of the XY model
(apart from a constant, the identification with Eq.~(\ref{eq:cXY}) takes place
by putting $\mu=2\lambda$), while for $r=L-1$ one gets the fully coordinated
fermionic system.
It is worth noticing that the spin Hamiltonian obtained by the inverse
Jordan-Wigner transformation of Eq.~(\ref{eq:fevrcXY}) is not the variable range
extension of the XY model \cite{keating04}.
For example, the fully coordinated spin system, known
as the Lipkin-Meshkov-Glick model \cite{lmg}, does not correspond to the fully coordinated
fermionic graph discussed below.
We also point out that, apart from a constant, the sign of the Hamiltonian is
changed by the canonical transformation $c_j'=c_j^\dagger$, so that the GS
phase diagram of $-H$ is equal to that of $H$. This is why both the
ferromagnetic and antiferromagnetic coupling in the XY-model give rise to the
same phase diagram (except for the sign of the magnetic field).

It is easy to express our variable range Hamiltonian in the form of
Eq.~(\ref{eq:qfH}). Since in the Hamiltonian the couplings only depend on the
distance between sites, $A$ and $B$ are Toeplitz matrices, i.e., the element of
the $j$-th row and the $k$-th column depends only on the difference $j-k$.
In the free-ends case, by using the discrete step function $\theta$ one can
write
\begin{eqnarray}
A_{jk}(r) =
-[(\mu-1)\delta_{jk}+\theta(r-|j-k|)] \ , \\
B_{jk}(r) =
-\gamma\,\mathrm{sign}(k-j)\theta(r-|j-k|) \ ,
\end{eqnarray}
where $j,k=1,\dots,L$, $r=0,\dots,L-1$, and we adopted the usual conventions
$\theta(0)=1$, $\mathrm{sign}(0)=0$.
In the case of periodic boundary conditions, where translational invariance
modulo $L$ takes place, $A$ and $B$ become circulant, i.e., each row is a
circular shift of the previous one.
For example $A_{21}=A_{1L}$, corresponding to the fact that the weight of the
term $c_2^\dagger{}c_1^{}$ in the Hamiltonian must be equal to that of the term
$c_{L+1}^\dagger{}c_L^{}=c_1^\dagger{}c_L^{}$.
The matrices can then be written as
\begin{eqnarray}
A_{jk}(r)
& = & -[(\mu-1)\delta_{jk}+\theta(r-|j-k|)+ \nonumber\\
&& +\theta(|j-k|-L+r)] \ , \\
B_{jk}(r)
& = & -\gamma\,\mathrm{sign}(k-j)[\theta(r-|j-k|)+
\nonumber\\
&& -\theta(|j-k|-L+r)] \ ,
\end{eqnarray}
where now $r=0,\dots,\lfloor(L-1)/2\rfloor$, with $\lfloor\dots\rfloor$
denoting the `floor' of a real number. The cyclic case $r=L/2$ for $L$ even
(corresponding to the fully coordinated system) has
to be treated separately: in this case $A_{jk}=-[(\mu-1)\delta_{jk}+1$, while
$B_{jk}=B_{jk}(r=L/2-1)$, i.e., $B_{1,L/2+1}=0$
\footnote{This is due to the anticommutation rules of fermionic operators: for
$L$ even $c_1^\dagger{}c_{L/2+1}^\dagger$ must have the same weight as
$c_{L/2+1}^\dagger{}c_{L+1}^\dagger=c_{L/2+1}^\dagger{}c_1^\dagger$, because of
periodicity, but this in turn is nothing but $-c_1^\dagger{}c_{L/2+1}^\dagger$,
and the two terms cancel in the Hamiltonian. Then, the corresponding entries in
the antisymmetric matrix $B$ have to vanish.}.

In order to visualize at a glance the simple structure of the above
matrices, we write explicitly the matrix $B$ for the case $r=2$ with periodic
boundary conditions, which is
\be
\rule{0cm}{2.5cm}B = -\gamma\left(\begin{array}{r*{6}{c}r}
\rule{0cm}{0.5cm}0 & 1 & 1 & 0 & \dots & 0 & -1 & -1 \\
\rule{0cm}{0.5cm}-1 & \ddots & \ddots & \ddots & \ddots && \ddots & -1 \\
\rule{0cm}{0.5cm}-1 & \ddots & \ddots & \ddots & \ddots & \ddots && 0 \\
\rule{0cm}{0.5cm}0 & \ddots & \ddots & \ddots & \ddots & \ddots & \ddots &
\vdots \\
\rule{0cm}{0.5cm}\vdots & \ddots & \ddots & \ddots & \ddots & \ddots & \ddots & 0 \\
\rule{0cm}{0.5cm}0 && \ddots & \ddots & \ddots & \ddots & \ddots & 1 \\
\rule{0cm}{0.5cm}1 & \ddots && \ddots & \ddots & \ddots & \ddots & 1 \\
\rule{0cm}{0.5cm}1 & 1 & 0 & \dots & 0 & -1 & -1 & 0
\end{array}\right) \ .
\ee

We discuss in some detail the cyclic case. Circulant matrices have some useful
properties which make their analysis very simple: (i) they all can be
diagonalized by the same unitary transformation and (ii) their eigenvalues are
given by the discrete Fourier transform of their first row.
Property (i) implies that circulant matrices commute and that sums and products
of circulant matrices are also circulant (as the transpose as well).
Furthermore, whenever a circulant matrix is invertible, the inverse is also
circulant. In brief, circulant matrices form a commutative ring.
The diagonalizing unitary matrix is usually written as
\be
\label{eq:circ_unitary}
\psi_{jk}^{\mathrm{(circ)}} = \frac{1}{\sqrt{L}}e^{-i2\pi(j-1)(k-1)/L} \ ,
\ee
where $j,k=1,\dots,L$.
The eigenvalues, according to property (ii), are instead given by
\be
\label{eq:circ_eigvals}
\lambda_{j}^{\mathrm{(circ)}} = \sum_{k=1}^Lm_{1k}e^{-i2\pi(j-1)(k-1)/L} \ ,
\ee
where $m_{jk}$ are the entries of the circulant matrix and again
$j,k=1,\dots,L$.
Note that $\lambda_1^{\mathrm{(circ)}}=\sum_{k=1}^Lm_{1k}$, while, in the case
of real matrices,
$\lambda_{L-j+2}^{\mathrm{(circ)}}=\lambda_j^{\mathrm{(circ)*}}$ for
$j=2,\dots,L$, where the star denotes complex conjugation.
Consequently, whenever the circulant matrix is real, one has
$\lfloor(L-1)/2\rfloor$ pairs of complex conjugate eigenvalues, with the
additional unpaired real eigenvalue $\lambda_1^{\mathrm{(circ)}}$ for $L$ odd
and the two additional unpaired real eigenvalues
$\lambda_1^{\mathrm{(circ)}},\lambda_{L/2+1}^{\mathrm{(circ)}}$ for $L$ even.
For symmetric matrices, where eigenvalues are real, complex conjugate pairs
become degenerate, so that the columns of matrix (\ref{eq:circ_unitary}) can be
linearly combined in pairs to get real eigenvectors.

The diagonalization of $A$, $B$, $Z=A-B$ is then straightforward. We denote by
$\alpha_j\in\R$, $\beta_j\in{}i\R$, $\zeta_j=\alpha_j-\beta_j$
the eigenvalues of $A$, $B$, $Z$ respectively ($j=1,\dots,L$). Since $Z$ is real,
$\zeta\in\mathrm{Sp}(Z)\Rightarrow\zeta^*\in\mathrm{Sp}(Z)$.
Clearly, in the polar decomposition $Z=\Lambda_\Phi{}T$ the positive part
$\Lambda_\Phi=\sqrt{ZZ^T}=|Z|$ and the unitary part $T$ are also circulant.
If $Z$ is invertible, $T=Z/|Z|^{-1}$ and its eigenvalues are simply given by
$\tau_j=e^{i\theta_j}=\zeta_j/|\zeta_j|$, i.e.,
$\theta_j=\arg(\zeta_j)=\arccos(\Re\zeta_j/|\zeta_j|)$.
Then from Eq.~(\ref{overlap-marco}) one gets
\be
\mathcal{F}(Z,\tilde{Z}) =
\sqrt{\prod_{j=1}^{L} \frac{1+e^{i(\tilde\theta_j-\theta_j)}}{2}} =
\delta_\Theta\prod_{j=2}^{M+1} \left|\cos\frac{\Theta_j}{2}\right| \ ,
\label{eq:Fcyclic}
\ee
where $\Theta_j=\tilde\theta_j-\theta_j$, $M=\lfloor(L-1)/2\rfloor$ is the
number of complex conjugate pairs, and $\delta_\Theta$ is a function which is
zero or one depending on whether $\Theta_1$ (and $\Theta_{L/2+1}$ for $L$ even)
is $\pi$ or zero (modulo $2\pi$) respectively.
Hence, either all of the unpaired $L-2M$ eigenvalues of $T^{-1}\tilde{T}$ are
equal to one, or the fidelity vanishes, corresponding to a
parity change in the ground state.

We evaluate explicitly the eigenvalues for the variable range $c$-cyclic case.
By using Eq.~(\ref{eq:circ_eigvals}) one finds
\begin{eqnarray}
\label{eq:zeta(r)}
-\zeta_j(r) & = &
\mu+(-1)^j\delta_{r,\lfloor(L+1)/2\rfloor}+ \\
&&+2\sum_{k=1}^r\left[\cos\frac{2\pi{}k(j-1)}{L}+
i\gamma\sin\frac{2\pi{}k(j-1)}{L}\right] \ , \nonumber
\end{eqnarray}
where $r=1,\dots,\lceil(L-1)/2\rceil$, with $\lceil\dots\rceil$ denoting the
`ceiling' of a real number.
The function $\delta_{r,\lfloor(L+1)/2\rfloor}$ is non-zero only for $r=L/2$,
with $L$ even.
The sums in Eq.~(\ref{eq:zeta(r)}) can be
usefully rewritten as
$\sum_{k=1}^r\cos(k\theta)=
\sin(r\theta/2)\cos[(1+r)\theta/2]/\sin(\theta/2)$ and
$\sum_{k=1}^r\sin(k\theta)=
\sin(r\theta/2)\sin[(1+r)\theta/2]/\sin(\theta/2)$.
The eigenvalues of the matrix $Z$ then assume the following expression
\begin{eqnarray}
\label{eq:zeta(r)summed}
-\zeta_j(r) & = &
\mu+(-1)^j\delta_{r,\lfloor(L+1)/2\rfloor}+ \\
&& +2\frac{\sin(r\kappa_j/2)}{\sin(\kappa_j/2)}
\left[\cos\!\left(\!\frac{1+r}{2}\kappa_j\!\right)\!+\!
i\gamma\sin\!\left(\!\frac{1+r}{2}\kappa_j\!\right)\right]
, \nonumber
\end{eqnarray}
where $\kappa_j=2\pi(j-1)/L$ and again $j=1,\dots,L$.
For the nearest neighbor range $r=1$ one recovers the well-known result of the
XY-model, which, in terms of the single particle spectrum, yields
$\Lambda_j(r=1)=|\zeta_j(r=1)|=
2\sqrt{(\cos\kappa_j+\mu/2)^2+\gamma^2\sin^2\kappa_j}$
\footnote{Recall that $\mu=2\lambda$ in terms of the
Hamiltonian~(\ref{eq:XY}).}.

\section{Fully coordinated fermionic Hamiltonian}

In this section we discuss the long range fermionic system, where, apart from
the diagonal elements, $A$ is the adjacency matrix of the
complete graph. The entries of the matrix $B$ can instead be interpreted as the
orientations of the couplings, so that we have the structure of a directed
graph.
To be consistent with the notation of Ref.~\cite{za-co-gio} in the following we
change the sign of the Hamiltonian with respect to the previous section, so that
$A\to-A$, $B\to-B$, $Z\to-Z$.
As noticed above, this does not affect the phase diagram.

\subsection{Cyclic case}
\label{subsec:cyclic}

We proceed by considering the cyclic fully coordinated case
$r=\lceil(L-1)/2\rceil$.
One then finds $\zeta_1=\mu+L-1$ and
\begin{eqnarray}
\label{eq:zeta_even}
\zeta_j & = & \mu-1+i\gamma[1+(-1)^j]\cot\frac{\kappa_j}{2}
\quad (L\ \mathrm{even}) \ , \\
\label{eq:zeta_odd}
\zeta_j & = &
\mu-1+i\gamma\left[\frac{1+(-1)^j}{2}\cot\frac{\kappa_j}{4}+ \right.\nonumber \\
&& \left.-\frac{1+(-1)^{j+1}}{2}\tan\frac{\kappa_j}{4}\right]
\quad (L\ \mathrm{odd}) \ ,
\end{eqnarray}
for $j=2,\dots,L$, where $\kappa_j=2\pi(j-1)/L$.
Hence, concerning the imaginary part $\Im\zeta_j=i\beta_j$, for $L$ even one has
$\Im\zeta_{2k}=2\gamma\cot[\pi(2k-1)/L]$ and $\Im\zeta_{2k-1}=0$, while for $L$
odd $\Im\zeta_{2k}=\gamma\cot[\pi(2k-1)/2L]$ and
$\Im\zeta_{2k-1}=-\gamma\tan[\pi(2k-2)/2L]$.
For the real part $\Re\zeta_j=\alpha_j$, we note that the above simple result
can also be readily obtained by considering the expression
\begin{eqnarray}
\label{eq:Aproj}
A & = & L\ket{\psi_1}\bra{\psi_1}+
(\mu-1)\openone = \\
& = & (L+\mu-1)\ket{\psi_1}\bra{\psi_1}+
(\mu-1)(\openone-\ket{\psi_1}\bra{\psi_1}) \ , \nonumber
\end{eqnarray}
where $\ket{\psi_1}$ is the uniform vector given by the first column of the
matrix $\psi^{\mathrm{circ}}$ defined in Eq.~(\ref{eq:circ_unitary}), in
components $\psi_{j1}^{\mathrm{(circ)}}=1/\sqrt{L}$.
Equation~(\ref{eq:Aproj}) directly diagonalizes $A$, i.e.,
$\textrm{Sp}A=\{L+\mu-1,\mu-1\}$, where the eigenvalue $\alpha=\mu-1$ is
$L-1$ times degenerate.

To proceed with our analysis we now fix $L$ odd, as usual for the XY-model.
We express the fidelity in the product form of Eq.~(\ref{eq:Fcyclic}) and
calculate the derivatives entering Eq.~(\ref{eq:h(Z)}).
We neglect the function $\delta_\Theta$ in Eq.~(\ref{eq:Fcyclic}), as it only
gives rise to the first order QPT placed at $\mu=1-L$, which is uninteresting
in the thermodynamic limit $L\to\infty$.
Then
\be
\left.
\frac{\partial^2F_Z}{\partial\lambda_k\partial\lambda_l}
\right|_{\lambda_h=0} =
-\frac{1}{4}\sum_{j=2}^{M+1}\left.\frac{\partial\Theta_j}{\partial\lambda_k}\right|_{\lambda_h=0}
\left.\frac{\partial\Theta_j}{\partial\lambda_l}\right|_{\lambda_h=0} \ ,
\ee
where the parameters $\{\lambda_h\}_{h=1,2}$ are the variations
$\lambda_1=\delta\mu$, $\lambda_2=\delta\gamma$ and one has
\begin{subequations}
\begin{eqnarray}
\partial_{\delta\mu}^2F_Z & = &
-\frac{\gamma^2}{4}S  \ , \\
\partial_{\delta\gamma}^2F_Z & = &
-\frac{(\mu-1)^2}{4}S \ , \\
\partial_{\delta\mu}\partial_{\delta\gamma}F_Z & = &
\frac{(\mu-1)\gamma}{4}S \ ,
\end{eqnarray}
\end{subequations}
with $S=\sum_{j=2}^{M+1}s_j$ and
$s_j\equiv(\partial_\gamma\Im\zeta_j)^2/|\zeta_j|^4$.
Consequently, one straightforwardly finds
\be
h[Z(\mu,\gamma)] = -\frac{1}{4}[(\mu-1)^2+\gamma^2]S \ ,
\ee
while the other Hessian eigenvalue is always zero and hence
$\det{H_{F_Z}(0)}=0$.
We also note that, apart from the discussion on first order QPTs, all these
results apply equally well to the even $L$ case.

We are interested in studying $h$ in the limit $L\to\infty$, where a
divergence of $h$ reflects a fast fidelity drop.
We therefore analyze the behavior of the functions $s_j$ entering the sum
$S$.
For $\mu\neq1,\gamma\neq0$ none of these functions is singular, independently of
$L$, so that in the thermodynamic limit one can approximate their sum with an
integral
\begin{eqnarray}
S(\mu\neq1,\gamma\neq0) & \sim & \frac{L}{\pi}\int_0^{\pi/4}
\left\{
\frac{\tan^2{x}}{[(\mu-1)^2+(\gamma\tan{x})^2]^2}+ \right.\nonumber\\
&&\left.+\frac{\cot^2{x}}{[(\mu-1)^2+(\gamma\cot{x})^2]^2}
\right\}\,\mathrm{d}x = \nonumber\\
& = & \frac{1}{4}\frac{L}{|\mu-1||\gamma|(|\mu-1|+|\gamma|)^2} \ ,
\end{eqnarray}
from which it follows the divergence $h\propto{}L$.
This corresponds to the fact that, for \textit{any} value of the Hamiltonian
parameters, the fidelity between different ground states vanishes in the
thermodynamic limit.

A different situation takes place for $\mu=1,\gamma\neq0$.
In this case one has
$s_{2j+1}=\cot^2(\pi{}j/L)/\gamma^4\sim{}L^2/(\pi{}j\gamma^2)^2$ for
$L\to\infty$, so that it is readily seen that the sum $S$ scales at least as
$L^2$.
For large $L$ we find
\bae
S & = & \frac{1}{\gamma^4}\sum_{j=1}^{O(L)}
\cot^2\frac{\pi{}j}{L}+O(L) = \nonumber \\
& = & \frac{1}{\gamma^4}\sum_{j=1}^{O(L)}
\left(\frac{L}{\pi{}j}\right)^2+O(L) \sim \frac{1}{6}\frac{L^2}{\gamma^4} \ ,
\eae
were in the last summation we used $O(L)\to\infty$ and hence the well known
result for the Riemann $\zeta$ function
$\zeta(2)=\sum_{j=1}^\infty{}j^{-2}=\pi^2/6$.
One then finds that the line $\mu=1$ corresponds to a faster decay of the
fidelity in the limit $L\to\infty$. This puts in evidence the presence of a
second order QPT, similarly to the case of the XY-model \cite{za-pa} for
$\mu=2$ (i.e., $\lambda=1$).
The same analysis can be repeated for $\mu\neq1,\gamma=0$, where
we get $S\sim(1/2)[L/(\mu-1)^2]^2$.
Summarizing, one has
\begin{subequations}
\label{eq:h_asymp}
\begin{eqnarray}
\label{eq:h_nocr_cyc}
h(\mu\neq1,\gamma\neq0) & \sim &
-\frac{L}{16}\frac{(\mu-1)^2+\gamma^2}
{|\mu-1||\gamma|(|\mu-1|+|\gamma|)^2} \ , \qquad \\
\label{eq:h_mu1_cyc}
h(\mu=1,\gamma\neq0)
& \sim & -\frac{1}{24}\left(\frac{L}{\gamma}\right)^2 \ , \\
\label{eq:h_gam0_cyc}
h(\mu\neq1,\gamma=0)
& \sim & -\frac{1}{8}\left(\frac{L}{\mu-1}\right)^2 \ .
\end{eqnarray}
\end{subequations}

\begin{figure}[t]
\includegraphics[width=8.5cm]{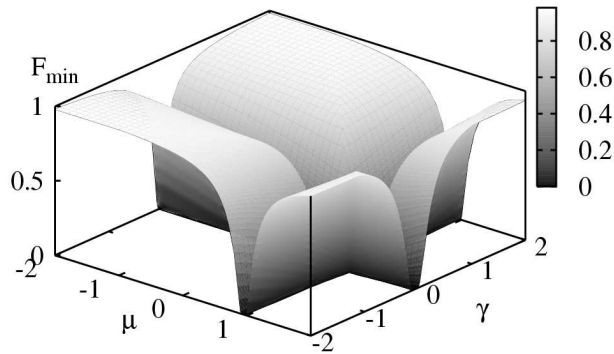}
\caption{\label{fig:F_min_cyc}%
Fidelity $\mathcal{F}_{\text{min}}=\min[\mathcal{F}(Z,\tilde{Z}_{\delta\mu}),
\mathcal{F}(Z,\tilde{Z}_{\delta\gamma})]$ (see Ref.~\cite{za-co-gio})
in the $\mu$-$\gamma$ plane for $L=1001$, with $\delta\mu=\delta\gamma=0.1$. At
the critical lines $\mu=1$ and $\gamma=0$ one observes an evident drop.}
\end{figure}
\begin{figure}[t]
\includegraphics[width=8.5cm]{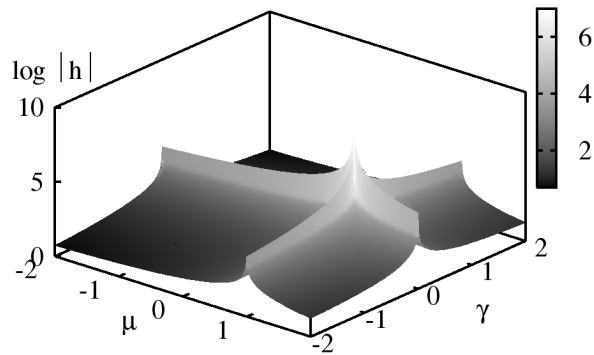}
\caption{\label{fig:h(Z)cyc}%
Decimal logarithm of the absolute value of $h(Z)$ in the $\mu$-$\gamma$ plane
for $L=1001$. Note the singular behavior at $\mu=1,\gamma=0$, where the
function is non-analytic. The critical lines $\mu=1$ and $\gamma=0$ emerge in a
clear way.}
\end{figure}

Finally, we note that the point $\mu=1,\gamma=0$ has a peculiar singular
behavior. There, all but one ($\zeta_1=L+\mu-1$) the eigenvalues of $Z$ vanish
and the unitary $T$ is undefined (as at a first order QPT where level crossing
takes place). Consequently, the angles $\Theta_j$ cannot be evaluated and the
above formalism does not apply.

The analysis of second order QPTs can independently be done in terms of the
derivatives of the GS energy
$E_0=\mathrm{Tr}(A-\Lambda)/2=\sum_{j=1}^L(\Re\zeta_j-|\zeta_j|)/2$.
Except on the $\mu$ axis, where $E_0=0$ for $\mu\geq1$ and $E_0=(L-1)(\mu-1)$
for $\mu<1$, the energy scales in a superextensive way, due to the fully
connected nature of the system.
Indeed, while $\mathrm{Tr}A=L\mu$, for $\gamma\neq0$ one has
$\mathrm{Tr}|B|\leq\mathrm{Tr}\Lambda\leq\mathrm{Tr}|A|+\mathrm{Tr}|B|\sim
\mathrm{Tr}|B|\propto{}L\ln{L}$
for $L\to\infty$.
This makes it impossible to define the energy density $E_0/L$ in the
thermodynamic limit.
The density of energy derivatives can however be calculated for finite $L$ and
in the TDL as well.
All the second derivatives are finite and continuous for $\mu\neq1$,
$\gamma\neq0$.
For $\mu\neq1$, $\gamma=0$ the second derivatives
$\partial_\gamma\partial_\mu{}E_0/L$ and $\partial_\gamma^2E_0/L$ are
analytical for $L$ finite, but in the TDL one finds the discontinuous behavior
$\partial_\gamma\partial_\mu{}E_0(\mu\neq1,\gamma\to0^\pm)/L\sim
\pm{}(1/\pi)/(\mu-1)$ and the divergence
$\partial_\gamma^2E_0(\mu\neq1,\gamma=0)/L\propto{}
S(\mu\neq1,\gamma=0)/L\propto{}L$.
It is hence reasonable to identify the critical line $\gamma=0$ with a
second order QPT.
Note, however, that the gap is finite at the critical point.
The discontinuity and the divergence of the second derivatives of energy and
fidelity at this gapful transition in the TDL is due to the divergence
$\propto{}L$ of some of the single particle energies $\Lambda_j=|\zeta_j|$.
This is a consequence of the unphysical long range nature of the fully
coordinated model.
Indeed, from Eq.~(\ref{eq:circ_eigvals}) one has
$|\zeta_j|\leq\sum_{k=1}^L|Z_{1k}|\leq\sum_{k=1}^L(|A_{1k}|+|B_{1k}|)$, so that
$|\zeta_j|\propto{}L$ only if $r\propto{}L$.
Another second order QPT is found at $\mu=1$, $\gamma\neq0$, where
$\partial_\mu^2E_0/L\propto{}\ln{L}$,
while the other second derivatives are continuous.
Here, however, one also has the vanishing of the lowest single particle energy,
since $\Re\zeta_j=0$ for $j=2,\dots,L$ and $\Im\zeta_k\to0$ for $L\to\infty$
with $k$ odd.

In Fig.~\ref{fig:F_min_cyc} we report the GS fidelity in the $\mu$-$\gamma$
plane, by using
$\mathcal{F}_{\text{min}}=\min[\mathcal{F}(Z,\tilde{Z}_{\delta\mu}),
\mathcal{F}(Z,\tilde{Z}_{\delta\gamma})]$, where
$\tilde{Z}_{\delta\mu}=Z(\mu+\delta\mu,\gamma)$ and
$\tilde{Z}_{\delta\gamma}=Z(\mu,\gamma+\delta\gamma)$, as in
Ref.~\cite{za-co-gio}.
This corresponds to plotting the minimum of the fidelity with respect to
variations along the two axes, which are related to $\partial_{\delta\mu}^2F_Z$
and $\partial_{\delta\gamma}^2F_Z$.
Since the critical lines are parallel to the axes, these variations are
sufficient to fully characterize the phase diagram.
A detailed analysis of all the second derivatives, including also
$\partial_{\delta\mu}\partial_{\delta\gamma}F_Z$, is instead provided by the study of
$h(Z)$.
The behavior of $h(Z)$ in the $\mu$-$\gamma$ plane is shown in
Fig.~\ref{fig:h(Z)cyc} for $L=1001$. The qualitative agreement with
Fig.~\ref{fig:F_min_cyc} is evident.

\begin{figure}[t]
\includegraphics[width=8.5cm]{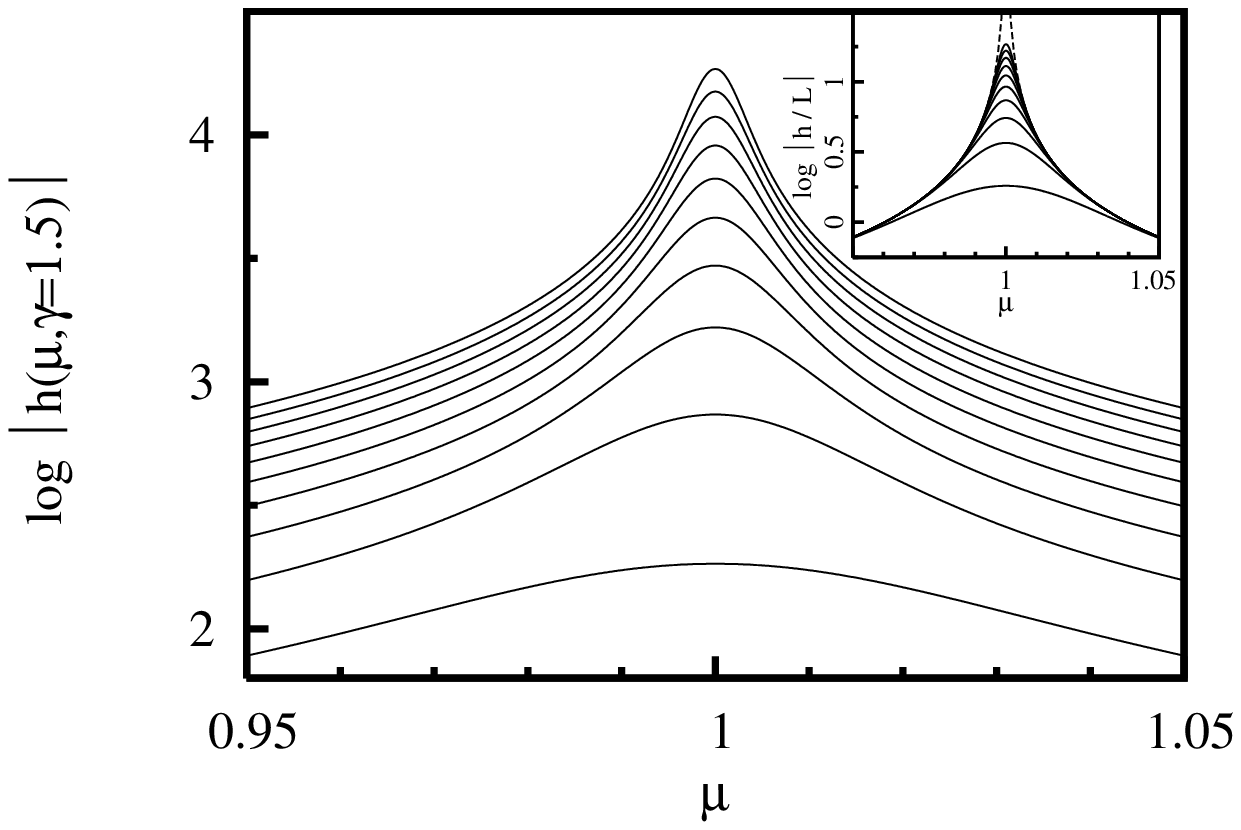}
\includegraphics[width=8.5cm]{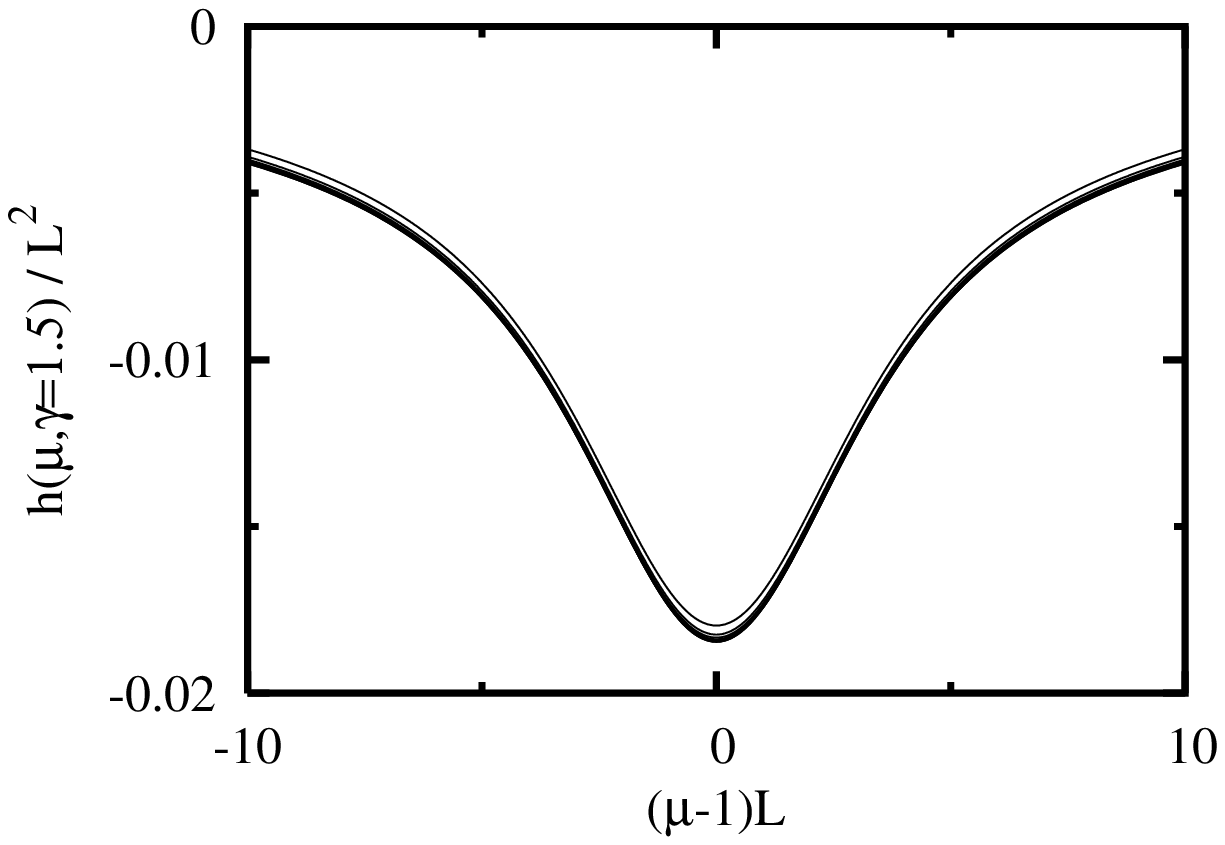}
\caption{\label{fig:cyc_fss}%
Finite-size scaling (upper panel) and data collapsing (lower panel) at the
critical line $\mu=1$. The upper figure shows the enhancement of the negative
peak of $h$ at $\mu=1$ for $\gamma=1.5$ and $L=101,201,\dots,1001$. In the
inset the curves are plotted after dividing $h$ by $L$, which allows a direct
comparison with the asymptotic curve (dashed line) given by
Eq.~(\ref{eq:h_nocr_cyc}).
The lower figure replots the same curves of the upper figure in rescaled units.
Most of them are practically indistinguishable and give rise to the thick line
in the plot.}
\end{figure}

Equations~(\ref{eq:h_asymp}) give the scaling of $h$ in the thermodynamic limit.
It is also interesting to look at the finite-size scaling of this function, in
connection with the next section about the free-ends fully coordinated system,
where the simple analytical results obtained above for $L\to\infty$
are not available.
The finite-size scaling for the critical line $\mu=1$ is shown in the upper
panel of Fig.~\ref{fig:cyc_fss}.
The peak of $h$ at $\mu=1$ increases (in modulus) with $L^2$, as expected from
Eq.~(\ref{eq:h_mu1_cyc}).
Away from the minimum, the function is instead proportional to $L$, as from
Eq.~(\ref{eq:h_nocr_cyc}).
Consequently, the different scaling behavior makes the peak shape
progressively narrower as $L$ increases.
It is possible to plot the function $h$ in rescaled units in order to put in
evidence these qualitative considerations.
By dividing the peak depth by $L^2$ and rescaling the distance $\mu-1$ from the
critical point by $L$ (therefore compensating the peak narrowing) one gets a
curve practically independent of $L$ (data collapsing), as found in the lower
panel of Fig.~\ref{fig:cyc_fss}.
This can be understood by looking at Eq.~(\ref{eq:h_nocr_cyc}) in the proximity
of the critical line $\mu=1$, i.e., for $|\mu-1|\ll\gamma$, which is well
verified for the parameters of Fig.~\ref{fig:cyc_fss}. Then one finds
$h\sim-L/16|\mu-1||\gamma|$, so that, for a given value of $\gamma$, the quantity
$h/L^2$ turns out to depend only on the combination $|\mu-1|L$.
Similar results were observed for the critical line $\gamma=0$.

We conclude the analysis of the cyclic fully coordinated system with a couple
of additional remarks.
As anticipated above, $\det{H_{F_Z}(0)}=0$, as one eigenvalue identically
vanishes.
The components of the eigenvector $\bm{v}=(v_\mu,v_\gamma)$ associated with
this zero eigenvalue are readily found to be related by
$v_\gamma=-(\partial^2_{\delta\mu}F_Z/
\partial_{\delta\mu}\partial_{\delta\gamma}F_Z)v_\mu$.
Hence, since
$-\partial^2_{\delta\mu}F_Z/\partial_{\delta\mu}\partial_{\delta\gamma}F_Z=
\gamma/(\mu-1)$, the eigenvector is directed in the radial direction from the
point $\mu=1,\gamma=0$. By recalling that the two eigenvectors of the symmetric
matrix $H_{F_Z}(0)$ are mutually orthogonal, one immediately concludes that the
direction of fastest GS variation is the azimuthal one with respect to this
shifted origin.
Finally, we comment on first order QPTs for $L$ even. In this case one has two
first order critical lines, namely $\mu=1-L$ and $\mu=1$, corresponding to the
vanishing of the eigenvalues $\zeta_1=L+\mu-1$ and $\zeta_{L/2+1}=\mu-1$ of
$Z$.

\subsection{Free-ends case}
\label{subsec:free-ends}

In the following we discuss the free-ends fermionic graph.
Some results on this system were already reported in Ref.~\cite{za-co-gio}. In
this case the relevant matrices are not circulant and the analytic results used
before are not available. In particular, it is worth stressing that the matrix
$Z$ in the free-ends case is in general not normal, i.e., $[Z,Z^T]\neq0$.
We therefore proceed by calculating the phase diagram numerically.

Let us first recall some analytical facts \cite{za-co-gio}.
For $\gamma=0$ the resulting number-conserving single-particle
Hamiltonian is the same as for the cyclic case.
For $(\mu=0,\gamma=1)$ the matrices $Z$ and $Z^\dagger=Z^T$ become instead
lower and upper triangular respectively. By explicit computation one finds
$(ZZ^T)_{ij}=4\min(L-i, L-j)$, which has the last column and row
identically vanishing. Accordingly $0\in\text{Sp}\,|Z(0,1)|\;\forall\;{L}$.
\footnote{At ($\mu=0,\gamma=1$) the matrix $ZZ^T$ can be diagonalized
analytically for any $L$.
The eigenvalues, apart from the lowest one which is always zero,
obey the formula
$1+\tan^2[j\pi/(2L-1)]$, $j=1,\dots,L-1$.
In the standard basis, the components of the normalized eigenvector
corresponding to the $j$-th non zero eigenvalue are
$2\sin[j\pi(2k-1)/(2L-1)]\cos(k\pi)/\sqrt{2L-1}$ for $k=1,\dots,L-1$ and zero
for $k=L$. The only non-zero component of the zero eigenvalue eigenvector is
instead the $L$-th one.}.
We also recall that changing the sign of $\gamma$ simply corresponds to
transforming $Z$ into $Z^T$. Hence, since
$\text{Sp}(ZZ^T)=\text{Sp}(Z^TZ)$, the matrix $\Lambda$ is
not affected by the transformation.
Then one can define $\Psi(-\gamma)=\Phi(\gamma)$ and consequently
$\Phi(-\gamma)=\Psi(\gamma)$, where the matrices $\Phi$ and $\Psi$ were
introduced in Sec.~\ref{sec:model}.
This implies that
$T(-\gamma)=\Phi^T(-\gamma)\Psi(-\gamma)=
[\Phi^T(\gamma)\Psi(\gamma)]^T=T^T(\gamma)$,
so that the overlap behavior is symmetric with respect to the $\gamma=0$ axis.

We start from the analysis of first order QPTs. Differently from the cyclic
case, the position of these transitions in the parameter space depends on $L$
in a non-trivial way, reaching however a simple configuration in the
thermodynamic limit \cite{za-co-gio}.
First order QPTs are given by level crossings which take place when the lowest
single particle energy (the gap) exactly vanishes, i.e., when
$\det\Lambda=|\det{Z}|=0$. Typically, at these points one also has a
discontinuous change in $\det{T}$, which jumps from $1$ to $-1$ or vice versa
(corresponding to a sign change in $\det{Z}$), implying
$\mathcal{F}=\sqrt{\det[(\openone+T^{-1}\tilde{T})/2]}=0$ whenever $T$ and
$\tilde{T}$ are calculated at opposite sides of the transition.
Strictly speaking, at the transition point where $\det{Z}=0$ the unitary $T$ is
instead undefined, similarly to the azimuthal angle of polar coordinates at the
origin.

In order to calculate the position of first order QPTs in the parameter space
one then has to solve the equation $\det{Z}=0$. The corresponding curve in the
$\mu$-$\gamma$ plane depends on $L$. Furthermore, its thermodynamic limit
depends on the parity of $L$.

The results for $L$ even were already reported in Ref.~\cite{za-co-gio}.
For $L=2$ the model is trivially solvable and one finds
$\det{Z}=\mu^2+\gamma^2-1$, so that the critical line is the circle of radius one
centered at the origin. For $L=4$ and $L=6$ the equation $\det{Z}=0$ is of second
and third degree in $\gamma^2$, respectively, so that relatively simple
analytical solutions are available.
One gets in this way a sequence of closed curves crossing the $\mu$ axis at
$\mu=1-L$ and $\mu=1$, while the $\gamma$ axis at $\gamma=\pm1$.
Numerically, in the limit $L\to\infty$ one finds the boundary
$\gamma=\pm1,\mu\leq1$ and $|\gamma|\leq1,\mu=1$.

\begin{figure}[t]
\includegraphics[width=4cm]{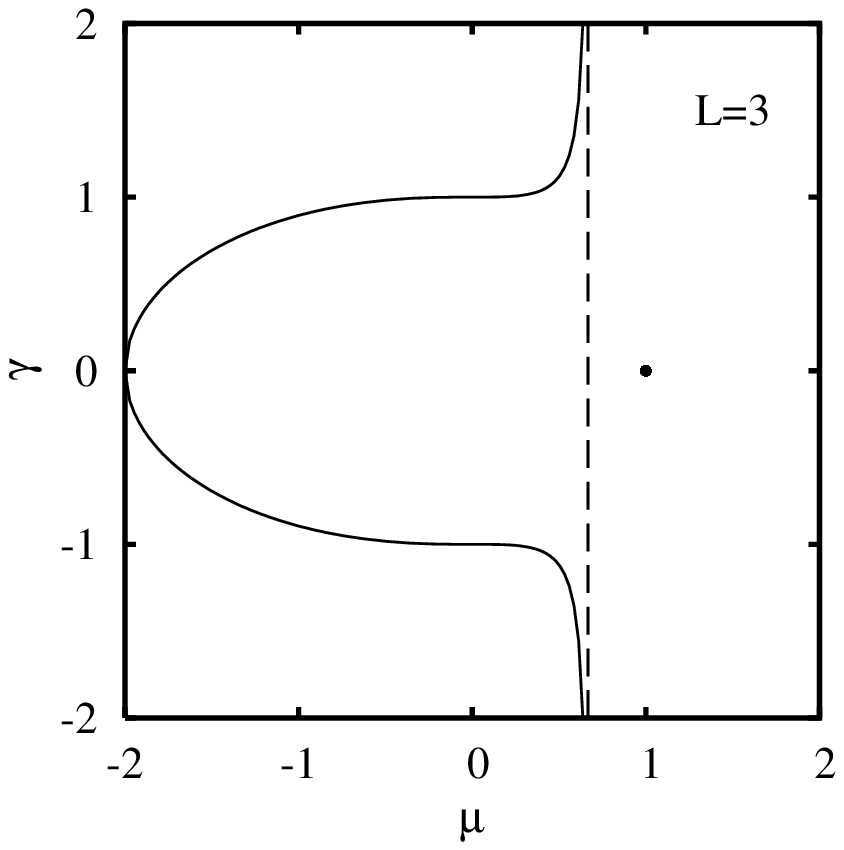}
\includegraphics[width=4cm]{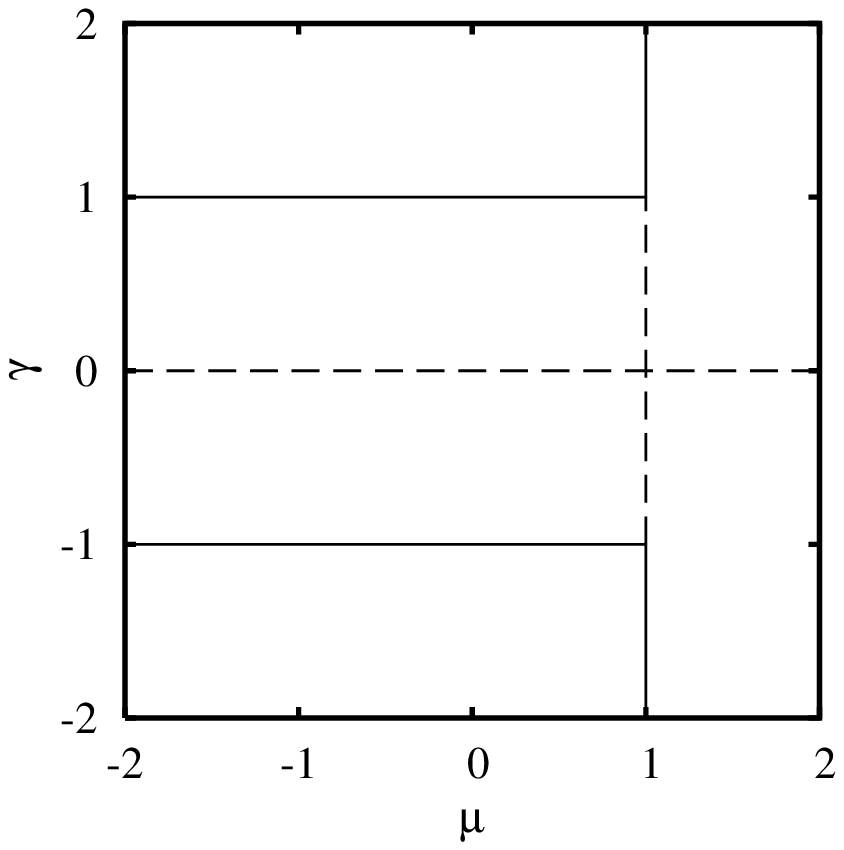}
\caption{\label{fig:detZ}%
Left panel: zeros of $\det{Z}$ in the $\mu$-$\gamma$ for $L=3$, given by
$\gamma^2=-(\mu-1)^2(\mu+2)/(3\mu-2)$.
Right panel: phase diagram in the $\mu$-$\gamma$ plane in the limit
$L\to\infty$ for $L$ odd. Solid and dashed lines correspond to first and second
order QPTs respectively.}
\end{figure}

For $L$ odd the results are different. The equation $\det{Z}=0$ for the case
$L=3$ is again readily solved analytically. For any $L$, the resulting curve is
now open and made of two branches. One branch starts at $\mu=1-L$, passes
through $\mu=0,\gamma=\pm1$, and diverges (as a function of $\mu$) at
$\mu=\mu_c$, where $\mu_c=2/3$ for $L=3$ (see left panel of
Fig.~\ref{fig:detZ}).
The second branch trivially reduces to the point $\mu=1,\gamma=0$.
The numerical analysis yields $\mu_c\to1$ in the thermodynamic limit, so that
the critical boundary (apart from the singular point $\mu=1,\gamma=0$) is given
by $\gamma=\pm1,\mu\leq1$ and $|\gamma|\geq1,\mu=1$
(see right panel of Fig.~\ref{fig:detZ}).

The fact that the asymptotic first order boundaries are different in the even
and odd $L$ cases is at first sight surprising. In the limit $L\to\infty$,
where $L+1\simeq{}L$, one would expect to find the same result in the two
cases.
However, the phases separated by the considered QPTs just differ by the fermion
number parity. The corresponding level crossings exchange energy states with
different parity (recall that the number parity is conserved by the
Hamiltonian).
Loosely speaking, these states, although orthogonal for any $L$, become less
and less different as $L$, and hence the average number of fermions contained
in the GS, increases.
In conclusion, in the thermodynamic limit such first order QPTs
\footnote{Similar first order QPTs happen also in the usual XY-model, as for
example at $\mu=-2$ (i.e., $\lambda=-1$) for $L$ odd in the $c$-cyclic case.}
are of little physical interest, in the sense that in practice it would be very
difficult to discriminate between such phases, as it would require the
capability to distinguish between $L$ and $L+1$ for $L\gg1$.
On the other hand, the different asymptotic behavior depending on the parity
of $L$ is not unreasonable for a quantity which is in turn related to a parity,
namely the fermion number parity.

We complete the analysis of the phase diagram by looking at the fidelity.
First order QPTs appear as sudden drops of the fidelity, which falls to zero
when the considered GSs are in different phases.
Besides these discontinuous drops, one also observes a smooth but evident
fidelity decay in the proximity of the lines $\mu=1$ and $\gamma=0$, which we
identify with second order transitions \cite{za-co-gio}.
Hence, apart from first order QPTs, the phase diagram of the free-ends case
corresponds to that of the cyclic case.

\begin{figure}[t]
\includegraphics[width=8.5cm]{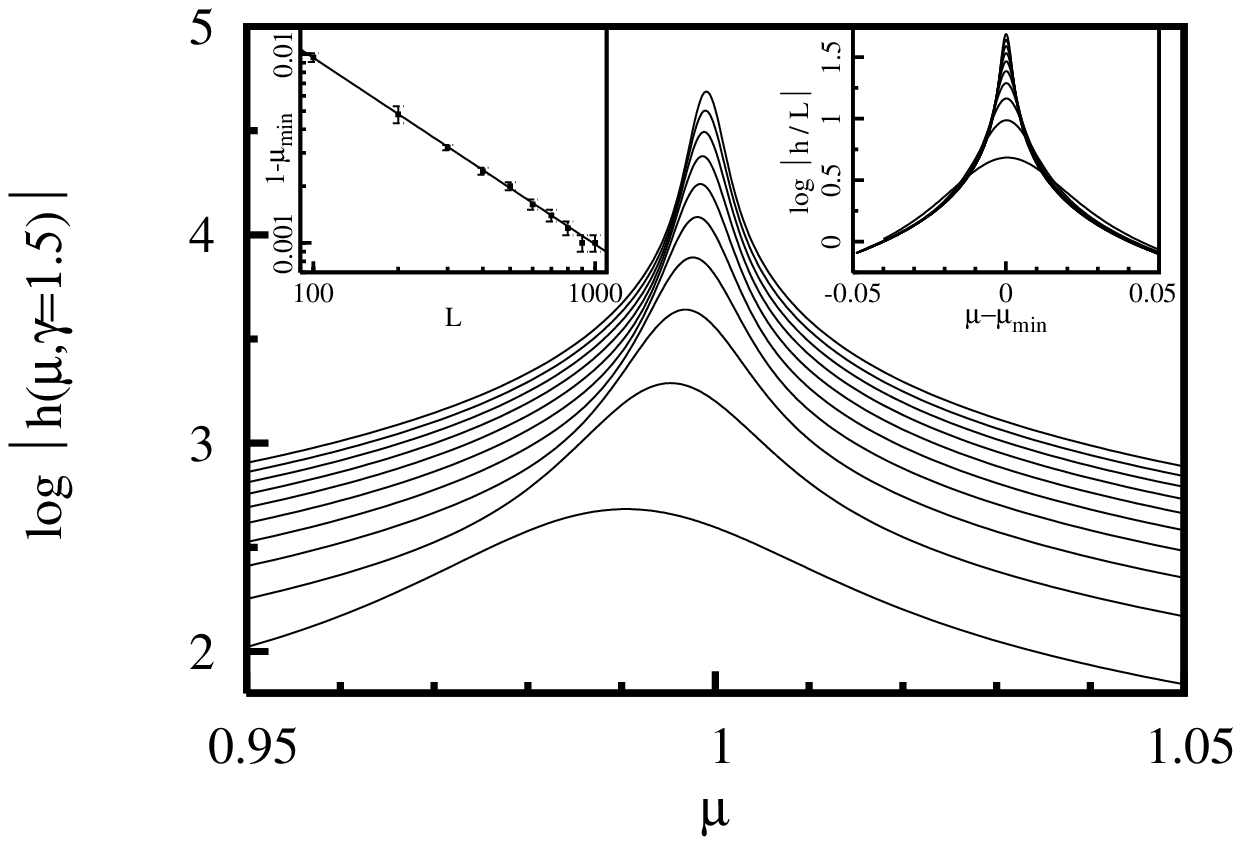}
\includegraphics[width=8.5cm]{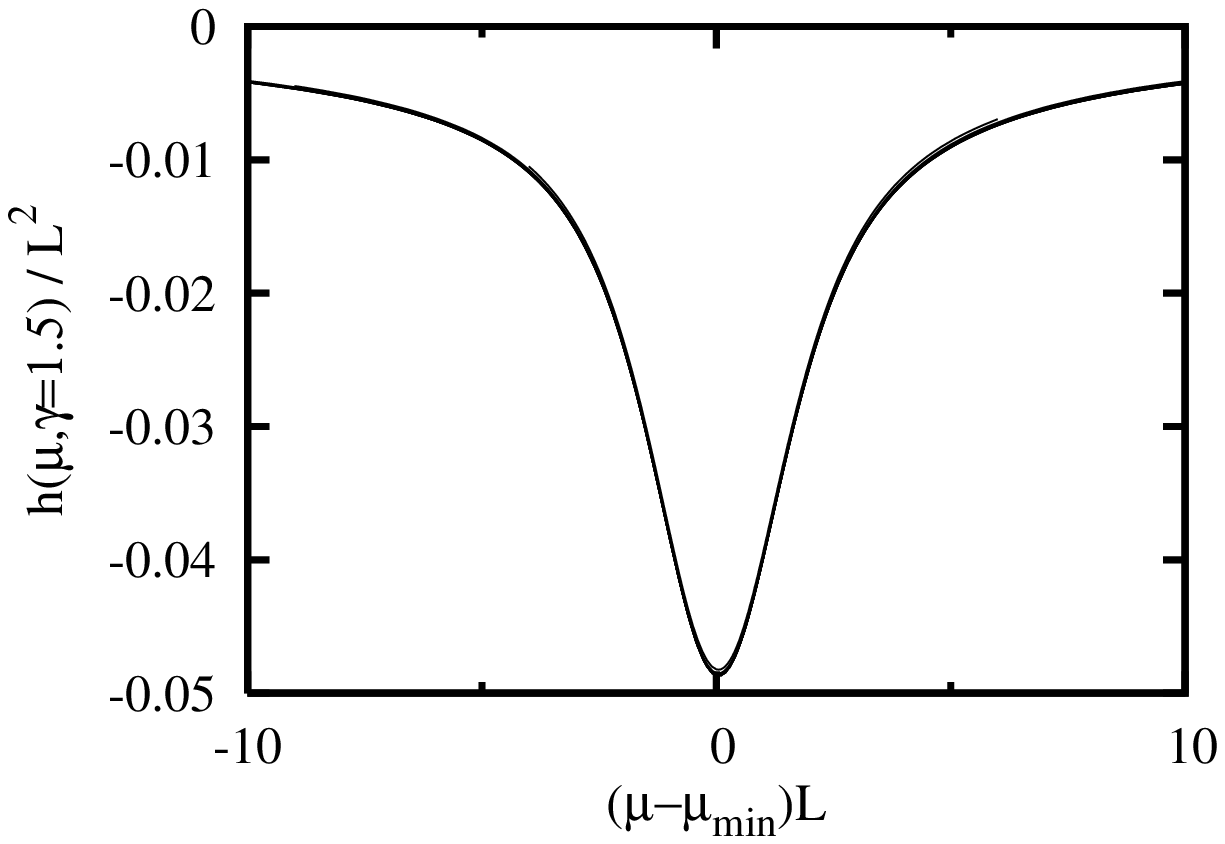}
\caption{\label{fig:fe_fss}%
Finite-size scaling (upper panel) and data collapsing (lower panel) at the
critical line $\mu=1$ for $\gamma>1$ and $L$ even, where (only) a second order
QPT is present.
The upper figure shows the enhancement of the peak of $h$ at $\mu=1$ for
$\gamma=1.5$ and $L=100,200,\dots,1000$.
In the insets, the law $1-\mu_{\mathrm{min}}\sim1/L$ for the peak position (error
bars correspond to numerical precision) and the function $h/L$ are shown.
In the lower figure,the same curves of the upper figure are replotted in rescaled
units, practically merging into an indistinguishable thick line as in the cyclic
case.}
\end{figure}

It is worth discussing in some detail the critical line $\mu=1$.
As explained above, for $L$ odd and $|\gamma|\geq1$ a first order QPT is also
present. For a given value $\gamma_0$ with $|\gamma_0|\in(1,\infty)$ and for
finite $L$ the (first order) transition point takes place at a critical
$\mu_{c}(\gamma_0)<1$ which tends to $1$ in the thermodynamic limit (see
Fig.~\ref{fig:detZ}).
At the point $\mu_{c}(\gamma_0),\gamma_0$ the fidelity function is
discontinuous and its derivatives, and hence the Hessian matrix, are not
defined. However, everywhere else on the line $\gamma=\gamma_0$ the eigenvalue
$h(Z)=\min\{\mathrm{Sp}[H_F(0)]\}$ can be calculated. This allows to analyze the
finite-size scaling behavior of $h$ in the neighborhood of $\mu=1$ even for
$|\gamma|\geq1$. One then finds the same divergence $\propto{}L^2$ as for second
order QPTs in the absence of first order transitions. Only, the constant of
proportionality is different in the two phases (i.e., for even or odd number of
fermions).
For $|\gamma|<1$, instead, one only has the second order transition and $h$ is
symmetric around the critical point. The position of the minimum of $h$ in the
thermodynamic limit gives the second order critical line. For finite $L$ this
position is given by the line $\mu=\mu_{\mathrm{min}}$, independently of
$\gamma$. In the thermodynamic limit the numerical analysis gives
$1-\mu_{\mathrm{min}}\sim{}1/L$, so that $\mu=1$ is the asymptotic critical
line.

For $L$ even, the first order QPT interesting the $\mu=1$ transition takes
place for $|\gamma|\leq1$. Except for this difference, everything goes as in
the odd $L$ case.
The finite-size scaling features of $h$ around $\mu=\mu_{\mathrm{min}}$ for $L$
even and $\gamma=1.5$ are summarized in Fig.~\ref{fig:fe_fss}.
It is also worth noticing that, albeit only in the TDL, the gap also vanishes
in the part of the $\mu=1$ critical line where the first order QPT is absent
(e.g., for $|\gamma|>1$ with $L$ even).

Another second order QPT is found along the line $\gamma=0$. Here the gap is
always finite, apart from the point $\mu=1$ (see the cyclic case). Nevertheless,
the finite-size scaling of $h$ puts in evidence the critical behavior. The
position of the minimum of $h$ is here independent of $L$, being always given by
the line $\gamma=0$.

Some insight in the nature of the latter transition can be gained by looking at
the perturbative expansion of the ground state around $\gamma=0$.
First of all we use a series expansion in $\gamma$ to prove analytically the
$L^2$ divergence of $h$ in $\gamma=0$.
Along the line $\gamma=0$, apart from the first order QPTs located at $\mu=1-L$
and $\mu=1$, the ground state is constant. Therefore
$\partial_{\delta\mu}^2F_Z=\partial_{\delta\mu}\partial_{\delta\gamma}F_Z=0$ and
consequently $h=\partial_{\delta\gamma}^2F_Z$.
In practice, one can reduce to the one-dimensional parameter space given by
$\gamma$ and use the series expansion in terms of ordinary derivatives presented
in \cite{za-co-gio},
$\mathcal{F}[Z(\lambda),Z(\lambda+\delta\lambda)]=
1+(1/16)\mathrm{Tr}(K'\delta\lambda)^2+O(\delta\lambda^3)$, where we assumed
$\det{T}=1$ so that $T=e^K$ and the prime denotes derivation with respect to
the one-dimensional parameter $\lambda$ (not to be confused with the magnetic
field strength in the XY-model).
For simplicity we consider the case $\mu>1$, where $T(\gamma=0)=\openone$ and
consequently $T'(\gamma=0)=T(\gamma=0)K'(\gamma=0)=K'(\gamma=0)$.
Lengthy but simple calculations (see Appendix~\ref{app:T'}) yield
\bae
\mathrm{Tr}[T'(\gamma=0)]^2 & = &
\frac{-1}{(\mu-1)^2}\frac{L(L-1)}{3[L+2(\mu-1)]^2}
[L^2+ \nonumber\\
&+&2L(2\mu-3)+4(\mu-1)(3\mu-5)] \ , \nonumber\\
\label{eq:Tr[T'(0)]}
\eae
so that one finds
$h(\gamma=0)=(1/8)\mathrm{Tr}[T'(\gamma=0)]^2\sim-(1/24)[L/(\mu-1)]^2$ in
the thermodynamic limit.
Again by series expansion one has $G(\gamma)=T'(\gamma=0)\gamma/2+O(\gamma^2)$.
The divergence of the elements of $T'$ at $\gamma=0$ (see
Appendix~\ref{app:T'}) in the thermodynamic limit therefore accounts for the
rapid orthogonalization rate around the critical line given by the $\mu$ axis.

\section{Single site entanglement}

The idea of systematically using measures of quantum correlations in order to
characterize QPTs has been explored in the past years for general models
\cite{entanglement} and in particular for the case of fermion systems
\cite{vidkit,fermionent}. It was found that some of the proposed
measures are able to spot, with their peculiar behavior, the undergoing QPTs;
in particular, in some cases the behavior of their derivatives is universal,
in the sense that they diverge with the correct critical exponents.
Here we would like to complete the fidelity analysis of QPTs in the cyclic
fully connected system of Subsec.~\ref{subsec:cyclic} by studying of one of the
simplest of such measures: the single site entanglement
$\mathcal{S}_i=- \mathrm{Tr}(\rho_i \log_2 \rho_i)$, where
$\rho_i=\sum_{k,h\in\{0,1\}} \rho_{h,k} \ketbra{h}{k} $ is the $2\times2$
single site reduced density matrix \cite{PZ02}.
It can be easily shown that for the system under study we have that
$\rho_{0,0}= 1-\rho_{1,1}=n$, where $n$ is the density.
Thus the single site entanglement is given by $\mathcal{S}_i=-n \log
n-(1-n)\log(1-n)$, where here the logarithms are taken in base $2$.
It can be explicitly computed by noticing that $n = \partial (E_0/L)/\partial
\mu$.
As already described, in  our case we have $E_0= (\mathrm{Tr} A -\mathrm{Tr}
\Lambda)/2=(L\mu-\sum_j |\zeta_j|)/2$ and $n$ can be obtained by calculating
the TDL of $\partial_\mu (\sum_j |\zeta_j| )/L = T_{ii}$. The last equality
follows from the following observations: (i) since $T$ is circulant all the
elements of its diagonal are equal and each of them corresponds to
$T_{ii}=(\mathrm{Tr} T)/L= \sum_j \tau_j/L$, where we recall that
$\tau_j=\zeta_j/|\zeta_j|$ are  the eigenvalues of $T$; (ii) if $\zeta_j \in
\mathrm{Sp}(Z)$ then $\zeta_j^* \in \mathrm{Sp}(Z)$ and thus $T_{ii}\in\R$;
(iii) the explicit formulas for $\zeta_j$ given in
Eqs.~(\ref{eq:zeta_even})-(\ref{eq:zeta_odd}) allow to write $\tau_1=1$, for
$\mu>1-L$, which is always true in the TDL, and $\partial_\mu |\zeta_j|=
(\mu-1)/|\zeta_j|=\Re \zeta_j /|\zeta_j|$ for $j=2,\dots,L$.
 We can thus write
 \be
   T_{ii}=\frac{1}{L}
   + \frac{1}{L}\sum_{j=2}^L \Re \zeta_j /|\zeta_j|=1-2n=
   \frac{\partial}{\partial \mu}\frac{\mathrm{Tr}\Lambda}{L} \ .
 \ee
The above derivation of $\mathcal{S}_i$ in terms of $T_{ii}$ shows the
physical link between the involved quantities and in particular with the
single particle eigenvalues $\zeta_j$. We would however like to remind that
the derivation could follow other lines of reasoning based on the well known
results about the evaluation of the entropy of entanglement
$\mathcal{S}_{l_1,l_2,\dots,l_N}$ of a block of $N$ sites
$\{l_1,l_2,\dots,l_N\}$.
In fact, in Refs.~\cite{vidkit,keating04}, it is shown how to express
$\mathcal{S}_{l_1,l_2,..,l_N}$ in terms of the eigenvalues $\nu_i$ of a matrix
$S_N=(T_N T_N^\dagger)^{1/2}$. There $T_N$ is the sub-block of order $N$,
relative to the $N$ sites belonging to the block, of a matrix (called $T$ in
Ref.~\cite{keating04}) equal to $\Psi^T\Phi$ and that is nothing but the transpose
of the $T$ used in this paper. The case of the single site entanglement is then
a special case of their analysis i.e., $N=1$ and, due to the circulant nature
of $T$ for all $i$ one has that $S_1=\nu_i=|T_{ii}|$ and
$\mathcal{S}_{l_i}=-[(1+\nu_i)/2] \log [(1+\nu_i)/2]-[(1-\nu_i)/2] \log
[(1-\nu_i)/2]$.

We now pass to the evaluation of the TDL of $T_{ii}$ for $L$ odd. We observe
that, since the single particle eigenvalues
$|\zeta_j|$ are different for $j$ even and odd,
we can split the sum in $T_{ii}$ into an even and an odd part and, since
in the TDL $\sum_k \rightarrow (L/\pi)\int$ , we can write these sums as
integrals such that:
\bae
T_{ii}&=&
\frac{1}{\pi}\int_0^{\pi/2}\frac{(\mu-1) \mathrm{d}x}{\sqrt{(\mu-1)^2+\gamma^2 \cot^2 x}} +
\nonumber \\
&+&\frac{1}{\pi}\int_0^{\pi/2}\frac{(\mu-1) \mathrm{d}x}{\sqrt{(\mu-1)^2+\gamma^2 \tan^2 x}}
\eae
The solution of these integral is different inside and outside the region
$|\gamma|>|\mu-1|$. One can write
\bae
T_{ii}^{'}&=&\frac{2}{\pi}
\frac{\mu-1}{\sqrt{\gamma^2-(\mu-1)^2}}
\ln \frac{\sqrt{\gamma^2-(\mu-1)^2}+|\gamma|}{|\mu-1|} \nonumber\\
\label{Tii'}
\eae
for $|\gamma|>|\mu-1|$ and
\be
T_{ii}^{''}=\frac{2}{\pi}
\frac{\mu-1}{\sqrt{(\mu-1)^2-\gamma^2}}
\arcsin{\frac{\sqrt{(\mu-1)^2-\gamma^2}}{|\mu-1|} }
\label{Tii''}
\ee
for $|\gamma|<|\mu-1|.$
$T_{ii}^{'}$ can be used to study the transition
$\mu \rightarrow 1$, while $T_{ii}^{''}$ will be used to study
the transition $\gamma \rightarrow 0$. As far as
$\mu \rightarrow 1$ is concerned, we first observe that
$T_{ii}^{'}\sim-(2/\pi)\ln|\mu-1|(\mu-1)/|\gamma|\rightarrow 0$, thus $n \rightarrow 1/2$
(half filling) and
$\mathcal{S}_i \rightarrow 1$ is maximal
On the contrary, when $\gamma \rightarrow 0$ we have that
$T_{ii}^{''} \rightarrow \mathrm{sign}(\mu-1)$, consequently $\mathcal{S}_i
\rightarrow 0$ and $n$ tends to zero or one, so that the ground state is
factorized.

The next step is the study of the derivatives of $T_{ii}$ and
$\mathcal{S}_i.$ One always has:
$\partial_x \mathcal{S}_i= (1/2)[\log((1-|T_{ii}|)/2) - \log((1-|T_{ii}|)/2)
]\partial_x |T_{ii}|$. We first analyze the case $x=\mu \ \  (\gamma \neq 0)$.
When $\mu \rightarrow 1$ we have that
$\partial_\mu T_{ii}'= -2\partial_\mu n = \partial^2_\mu E_0/L
\approx -\ln{|\mu-1|}/|\gamma|.$ 
The divergence of the density, that correctly signals the undergoing phase
transition, is however not transferred to the first derivative of the single site
entanglement; the latter has a maximum in $\mu=1$ and in fact we have
$\partial_\mu \mathcal{S}_i \propto \log{\frac{1-|T_{ii}'|}{1+|T_{ii}'|}}
\partial_\mu |T_{ii}'| \approx (\mu-1)\log^2{|\mu-1|}=0$. The divergence is
shifted to the derivative of second order, i.e., $\partial^2
\mathcal{S}_i/\partial \mu^2\approx -\log^2{|\mu-1|}$.

We now analyze the case $x=\gamma\ \  (\mu \neq 1)$. At the transition
$\gamma\to0$ the derivative
$\partial_\gamma T_{ii}''\to -\mbox{sign}(\gamma)2[\pi(\mu-1)]^{-1}$ is finite
but, since when $\gamma \rightarrow 0$, $T_{ii}'' \approx 1 -2 \gamma/(\pi
(\mu-1))$ the derivative of the single site entanglement
$\partial_\gamma\mathcal{S}_i$ diverges as
$-\mbox{sign}(\gamma)\log{(|\gamma|/\pi |\mu-1|)}/|\mu-1|$.

In summary we can make the following considerations.
Due to its direct link to the density $n$,
the behavior of $\mathcal{S}_i$ in the TDL reflects the properties of the ground state:
for $\gamma=0$ ($\mu \neq 1$) the state is factorized, while
for $\mu=1$ ($\gamma \neq 0$) the single site quantum correlations are
maximal. At the transitions the critical behavior is described
by the derivatives of $\mathcal{S}_i$. For $\mu \rightarrow 1$ the divergence
of $\partial^2 \mathcal{S}_i/\partial \mu^2$ is again directly linked to the
divergence of $\partial_\mu n$.
In the case $\gamma \rightarrow 0$, the derivative $\partial_\gamma T_{ii}''$
is finite and hence it is the functional form of the chosen measure of
entanglement $\mathcal{S}_i$ that is
responsible for the divergence. Indeed $\partial_\gamma\mathcal{S}_i \propto
-\log{n}+\log(1-n)$ and in this transition $n\rightarrow 0,1$ depending on
$\mathrm{sign}(\mu-1)$.

\section{Conclusions}

In this paper we have given an extensive discussion of the relation between
ground state fidelity and quantum phase transitions in quadratic Fermi
systems.
The presented material covers several aspects.
First (i) we have provided a detailed description of the ground state
calculation and a simple formula for the fidelity between ground states
corresponding to different parameters. The latter expression is based on the
orthogonal part $T$ of the polar decomposition of the coupling constant matrix
$Z$.
The fidelity behavior has then been characterized through a combination of its
second derivatives, namely the minimum eigenvalue $h$ of the Hessian matrix.
Subsequently (ii) we have introduced a class of models which encompass the XY
fermionic Hamiltonian and are analytically solvable in the cyclic case, thereby
offering a useful mean to exemplify the above concepts.
In particular, we have focused on the fully connected system, whose long range
nature has been shown to give rise to a peculiar gapful quantum phase
transition, where $h$ exhibits critical finite-size scaling properties.
Finally (iii) for the latter model in the cyclic case we have also provided an
analysis of the single site entanglement, which can be extracted from the same
matrix $T$ used to calculate the fidelity.

The fidelity approach to QPTs, besides being based on an intuitive
understanding of the ground state dramatic change in critical regions, seems to
provide further conceptual insight into these phenomena, making explicit the
connections among the ground state, its energy, and the single particle
spectrum.
In particular, the possibility of expressing the ground state of these systems
only as a function of the matrix $T$ completely clarifies its relation with the
single particle energies.
The latter are seen to give rise to divergences in fidelity derivatives when
some of them either tend to zero or to infinite in the thermodynamic limit.
The analytical solvable models presented here allow for a study of these
relations in a thorough way.

We deem it appropriate to add some remarks on the possible developments of this
approach.
In all the models considered so far, a one to one correspondence between QPTs
and fidelity drops has been observed. It would however be interesting to
study the fidelity behavior in the presence of more subtle transitions, as in
the case of topological QPTs, or even in crossover regions, where no
discontinuous transition is present.
In fact, it would be desirable to answer the question, whether fidelity drops
give a reliable signature of QPTs or if they can be originated also by to other
effects.
In other words, could fidelity drops (or, more precisely, divergences of
fidelity derivatives) provide a tool to \textit{define} QPTs?
In spite of the possible computational difficulty of the fidelity analysis in
general systems,
the added value due to the conceptual simplicity of the ideas underlying our
method makes the latter, we believe, an interesting question.

\appendix

\section{}
\label{app:parity}

In this appendix we prove the relationship between $\det{T}$ and the ground
state number parity.

We start by proving that $\det{T}=1$ implies even parity.
If $\det{T}=1$ and $-1\notin\mathrm{Sp}T$ then $T+\openone$ is invertible, $G$
exists, and the even parity follows immediately.
If instead $\det{T}=1$ but $-1\in\mathrm{Sp}T$, then this negative eigenvalue
must appear an even number of times in the spectrum.
We note that, if $\det{T}=1$, one can write $T=e^K$, where $K$ is an
antisymmetric real matrix.
Then, a real orthogonal transformation $R$ exists such that $R^TKR$, and hence
$R^TTR$, is block diagonal. For $L$ even, one has only $2\times2$ blocks, while
for $L$ odd an additional $1\times1$ block is present, corresponding to the
only unpaired eigenvalue, which is $0$ for $K$ and $1$ for $T$.
Being present an even number of $-1$ eigenvalues in $\mathrm{Sp}T$, one can
group them in pairs in blocks of the form $-\openone_2$.
The idea is now to find a parity preserving canonical transformation of the
$c_i$'s which flips the sign of the $-1$ eigenvalues of $T$. In this way one
gets a matrix $T'$ s.t. $\det{T}=1$ and $-1\notin\mathrm{Sp}T'$, being thus
possible to define $G'=(T'-\openone)/(T'+\openone)$. The ground state will then
have the form~(\ref{Psi}) with the transformed operators $c_i'$ and the
corresponding vacuum $\ket{0'}$, so that in terms of the new parity operator
$P_{N'}=(-1)^{N'}$, where $N'=\sum_{j=1}^L{c_j'}^\dagger{}c_j'$, the ground
state is even. The parity preserving nature of the canonical transformation
will ensure that $P_{N'}=P_N$ and the proposition be proved.

The transformation properties of the matrix $Z$ for a general real canonical
transformation have been described in the first part of Sec.~\ref{sec:model}.
We then consider the canonical transformation
$c_j'=R_{jk}^Tc_k$, i.e., $u=R^T$, $v=0$.
This transformation is evidently number preserving,
$N'=\sum_{j=1}^L{c_j'}^\dagger{}c_j'=\sum_{j=1}^Lc_j^\dagger{}c_j=N$.
Furthermore one has $Z'=R^TZR=R^T\sqrt{ZZ^T}RR^TTR$ so that $T'=R^TTR$ is block
diagonal.

Without loss of generality, we now assume that the first block of $T'$ is given
by $-\openone_2$ and that no other $-1$ appears in the spectrum.
We then use a second trivial canonical transformation defined by
$u'+v'=\openone$ and $u'-v'=-\openone_2\oplus\openone_{L-2}$.
In this way one gets $Z''=Z'(u'-v')$ and $T''$ is equal to $T'$ apart from the
first two diagonal elements, whose sign is now changed.
Explicitly, $u'=\mathbb{0}_2\oplus\openone_{L-2}$,
$v'=\openone_2\oplus\mathbb{0}_{L-2}$, and it is immediate to check that $u'$
and $v'$ satisfy the canonical conditions~(\ref{eq:can_cond}).
It is easy to see that this transformation, although not number conserving,
is parity preserving.
Indeed one finds $N''=N'+2(1-n_1'-n_2')$, where
$n_j'={c_j'}^\dagger{}c_j'$, so that $P_{N''}=P_{N'}=P_N$.

This proves that if $\det{T}=1$ the ground state is even.
Conversely, if $\det{T}=-1$, one can find a trivial parity flipping canonical
transformation which yields $T'$ such that $\det{T'}=1$. For example, one can
choose \cite{peschel} $c_1'=c_1^\dagger$ and $c_j'=c_j^{}$ for $j=2,\dots,L$,
i.e., $u=0\oplus\openone_{L-1}$ and $v=1\oplus\mathbb{0}_{L-1}$. Then
$u+v=\openone$ and $u-v=-1\oplus\openone_{L-1}$, so that
$T'=T(-1\oplus\openone_{L-1})$ and hence $\det{T'}=-\det{T}=1$. In addition,
$N'=N+1-2n_1$, so that the parity is flipped, $P_{N'}=-P_N$. The ground state
will then be of the form~(\ref{Psi}) with the transformed operators and vacuum
$\ket{0'}=c_1^\dagger\ket{0}$, being hence of even parity in the $c_j'$
representation and thus odd in the original one.
This completes our proof.

\section{}
\label{app:T'}

We briefly sketch the main steps of the calculation of
$T'(0)\equiv\partial_\gamma{}T|_{\gamma=0}$ used in Eq.~(\ref{eq:Tr[T'(0)]}) of
Subsec.~\ref{subsec:free-ends}. In the following, primed quantities imply
derivation with respect to $\gamma$.

By explicitly deriving $T=\Phi^T\Lambda^{-1}\Phi{}Z$ with respect to $\gamma$
and using the fact that $\Lambda'(0)=\mathbb{0}$ one finds
$T'(0)=[{\Phi'}^T(0)\Phi(0), A^{-1}]A-A^{-1}B'$, where $B'=B/\gamma$.

To avoid the explicit calculation of $\Phi'(0)$ one can use the relation
$[A,B']=[{\Phi'}^T(0)\Phi(0), A^2]$, obtained from $(ZZ^T)'(0)$ with the aid of
$\Lambda'(0)=\mathbb{0}$.
By noting that $A^2=[L+2(\mu-1)]A-(L+\mu-1)(\mu-1)\openone$ [see
Eq.~(\ref{eq:Aproj})]
and that $[{\Phi'}^T(0)\Phi(0), A^{-1}]A=-A^{-1}[{\Phi'}^T(0)\Phi(0), A]$ one
finds
\be
T'(0) = -A^{-1}\left(\frac{[A,B']}{L+2(\mu-1)}+B'\right) \ ,
\ee
which, after algebraic manipulations, yields
$T_{jk}'(0)=[(k-j)/(L/2+\mu-1)-\mathrm{sign}(k-j)]/(\mu-1)$.
Then, Eq.~(\ref{eq:Tr[T'(0)]}) is obtained by calculating only the diagonal
elements of $[T'(0)]^2$, namely
\bae
\{[T'(0)]^2\}_{jj} & = &
\frac{1}{(\mu-1)^2}\left[
\frac{2(\mu-1)j(j-L-1)}{(L/2+\mu-1)^2}+ \right.\nonumber\\
&+&\left.\frac{1}{6}\frac{L(L+1)(L+6\mu-2)}{(L/2+\mu-1)^2}-L+1\right] \ ,
\nonumber\\
\eae
and finally summing to get the trace.

\end{document}